\documentclass[12pt]{iopart}

\usepackage{subfigure}
\usepackage{graphics}
\usepackage{graphicx}
\begin{document}

\title{Transport of Self-propelled Janus Particles Confined in Corrugated Channel with L\'{e}vy Noise}

\author{Bing Wang}

\address{School of Mechanics and Optoelectronics Physics, Anhui University of Science and Technology, Huainan, 232001, P.R.China}
\ead{hnitwb@163.com}

\author{Zhongwei Qu}

\address{School of Mechanics and Optoelectronics Physics, Anhui University of Science and Technology, Huainan, 232001, P.R.China}

\author{Xuechao Li}

\address{School of Mechanics and Optoelectronics Physics, Anhui University of Science and Technology, Huainan, 232001, P.R.China}
\vspace{10pt}
\begin{indented}
\item[]August 2017
\end{indented}

\begin{abstract}
The transport of self-propelled particle confined in corrugated channel with L\'{e}vy noise is investigated. The parameters of L\'{e}vy noise(i.e., the stability index, the asymmetry parameter, the scale parameter, the location parameter) and the parameters of confined corrugated channel(i.e., the compartment length, the channel width and the bottleneck size) have joint effects on the system. There exits flow reverse phenomena with increasing mean parameter. Left distribution noise will induce $-x$ directional transport and right distribution noise will induce $+x$ directional transport. The distribution skewness will effect the moving direction of the particle. The average  velocity shows complex behavior with increasing stability index. The angle velocity and the angle Gaussian noise have little effects on the particle transport.
\end{abstract}

%
%
%
%
%

\section{\label{label1}Introduction}
Rectification of Brownian motion in a narrow, periodically corrugated channel has been the focus of a concerted effort aimed at establishing net particle transport in the absence of external biases. Some diffusive transport through microstructures is ubiquitous and attracts evergrowing attention from physicists\cite{Reguera2006,Lindenberg2007,Yang2017,Skaug2018,Bressloff,Bing}, engineers\cite{Berkowitz}, and biologists\cite{Hofling}. H\"{a}nggi \emph{et al}. presented an overview of artificial Brownian motors, attempted to explore future pathways and potential new applications of artificial Brownian motors\cite{Hanggi}. Brownian particles in regular arrays of rigid obstacles, and also in the corrugated geometry channel show many interesting phenomena.

Self-propelled particles performing directed motion by extracting energy from external environment, are rather different from traditional inertia particles which are dominated by thermal fluctuations. Self-propelled particle confined in channel has attracted widely attention\cite{Wu,Ao,Liu}. Ao \emph{et al}. investigated the transport diffusivity of Janus particles in the absence of external biases and found the self-diffusion constants depends on both the strength and the chirality of the self-propulsion mechanism\cite{Ao}. Ghosh \emph{et al}. investigated Brownian transport of self-propelled overdamped microswimmers in a two-dimensional periodically compartmentalized channel\cite{Ghosh}. Malgaretti \emph{et al}. analyzed the dynamics of Brownian ratchets in a confined environment and found the combined rectification mechanisms may lead to bidirectional transport\cite{Malgaretti}. Teeffelen \emph{et al}. studied the motion of a chiral swimmer in a confining channel and found self-propelled particles move along circles rather than along a straight line when their driving force does not coincide with their propagation direction\cite{Teeffelen}. Pototsky \emph{et al}. considered a colony of point like self-propelled surfactant particles without direct interactions that cover a thin liquid layer on a solid support\cite{Pototsky}.

All these studies devoted to the self-propelled particles were treating the input noises process as Gaussian noise. In practice, various non-Gaussian noises have distinct spiky and impulsive characteristics, the decay of its' probability density function is slower than the Gaussian distribution's, showing significant tails. The L{\'{e}}vy distribution, which bases on the generalized central limit theorem, has the statistical characteristics of non-Gaussian and heavy tailed. So it provides a strong theoretical tool for the analysis of the non-Gaussian noises and signals. L{\'{e}}vy noise frequently appears in areas of statistical mechanics, finance, and signal processing, is more suitable for modeling diversified system noise because it can be decomposed into a continuous part and a jump part by L{\'{e}}vy -It\^{o} decomposition\cite{Janicki,Applebaum2009,Applebaum2010,Di Nunno,Yuan}. L{\'{e}}vy noise extends Gaussian noise to many types of impulsive jump-noise processes found in real and model neurons as well as in models of finance and other random phenomena.

In this paper, we investigate the transport phenomenon of self-propelled Brownian particles confined in $2D$ corrugated channel with L\'{e}vy noise. The paper is organized as follows: In Section \ref{label2}, the basic model of self-propelled particles confined in a $2D$ channel with  L\'{e}vy noise is provided. In Section \ref{label3}, the effects of the channel and noise are investigated by means of simulations. In Section \ref{label4}, we get the conclusions.

\section{\label{label2}Basic model and methods}
In this work, we consider the self-propelled Brownian particles confined in a $2D$ sinusoidal channel. The dynamics of the particles can be described by the following Langevin equations\cite{Reguera2012}
\begin{equation}
\frac{dx}{dt}=v_0\cos\theta+\xi(t) \label{Ext}
\end{equation}
\begin{equation}
\frac{dy}{dt}=v_0\sin\theta+\xi(t) \label{Eyt}
\end{equation}
\begin{equation}
\frac{d\theta}{dt}=\omega+\xi_{\theta}(t) \label{Ethetat}
\end{equation}
$x$ and $y$ are the positions of the particle. $v_0$ is the self-propelled velocity. $\theta$ is the angle between the moving direction and the $x$ axis. $\omega$ is chosen so as to coincide respectively with the positive and negative chirality of the swimmer. $\xi(t)$ is the L\'{e}vy noise and obeys L\'{e}vy distribution $L_{\alpha, \beta}(\zeta; \sigma,\mu)$ , and the characteristic
function is\cite{Janicki}:
\begin{equation}
\Phi(k)=\int_{-\infty}^{+\infty}d{\zeta}\exp(ik\zeta)L_{\alpha,\beta}(\zeta;\sigma,\mu)
\end{equation}
for $\alpha\in(0,1)\cup(1,2]$
\begin{equation}
\Phi(k)=\exp\{i{\mu}k-\sigma^\alpha|k|^\alpha(1-i{\beta}{\rm{sgn}}(k)\tan\frac{\pi \alpha}{2})\}
\end{equation}
and for $\alpha=1$
\begin{equation}
\Phi(k)=\exp\{i{\mu}k-\sigma|k|(1+i{\beta}{\rm{sgn}}\frac{2}{\pi}\ln|k|)\}
\end{equation}

Here $\alpha\in(0,2]$ denotes the stability index that describes an asymptotic power law of the L\'{e}vy distribution. When $\alpha<2$, $L_{\alpha, \beta}(\zeta; \sigma,\mu)$  is characterized by a heavy-tail of $|\zeta|^{-(\alpha+1)}$ type with $|\zeta|\gg1$. The constant $\beta$  is the asymmetry parameter with $\beta\in[-1,1]$. When $\beta$ is positive, the distribution is skewed to the right. When it is negative, the distribution is skewed to the left. When $\beta=0$, the distribution is symmetrical. As $\alpha\rightarrow2$, the distribution approaches the symmetrical Gaussian distribution regardless of $\beta$. $\sigma$ is the scale parameter with $\sigma\in(0,\infty)$, $\mu(\mu\in{R})$ denotes the location parameter, and $D=\sigma^\alpha$ represents the noise intensity. In this paper we use the Janicki-Weron algorithm  to generate the L\'{e}vy distribution\cite{Janicki}.

As $\alpha\neq1$, $\xi$ is simulated as
\begin{equation}
\xi=D_{\alpha,\beta,\sigma}B_{\alpha, \beta}\{\frac{\cos(M-\alpha(M+C_{\alpha,\beta}))}{W}\}^{(1-\alpha)/\alpha}+\mu
\end{equation}

As $\alpha=1$, $\xi$ can be obtained from the formula
\begin{equation}
\xi=\sigma\frac{2}{\pi}\{(\frac{\pi}{2}+\beta{M})\tan(M)-\beta\ln(\frac{W\cos(M)}{\frac{\pi}{2}+\beta{M}})\}+\frac{2}{\pi}\beta\sigma{\ln\sigma}+\mu
\end{equation}
the constants $B_{\alpha, \beta}$, $C_{\alpha, \beta}$, $D_{\alpha,\beta,\sigma}$ are given by
\begin{equation}
B_{\alpha, \beta}=\frac{\sin(\alpha(M+C_{\alpha,\beta}))}{(\cos(M))^{1/\alpha}}
\end{equation}
\begin{equation}
C_{\alpha, \beta}=\frac{\arctan(\beta\tan(\frac{\pi\alpha}{2}))}{\alpha}
\end{equation}
\begin{equation}
D_{\alpha,\beta,\sigma}=\sigma\{1+\beta^2\tan^2(\frac{\pi\alpha}{2})\}^{1/2\alpha}
\end{equation}

$M$ is a random variable uniformly distributed over $(-\frac{\pi}{2},\frac{\pi}{2})$. $W$ is a random variable exponentially distributed with a unit mean. $M$ and $W$ are statistically independent\cite{Janicki,Weron,West}.

$\xi_\theta$ is the self-propelled angle Gaussian color noise, and describes the nonequilibrium angular fluctuation. $\xi_\theta$ satisfies the following relations
\begin{equation}
\langle\xi_\theta(t)\rangle=0
\end{equation}
\begin{equation}
\langle\xi_\theta(t)\xi_\theta(t')\rangle=\frac{Q_\theta}{\tau_\theta}\exp(-\frac{|t-t'|}{\tau_\theta})
\end{equation}
$\langle\cdots\rangle$ denotes an ensemble average over the distribution of the random forces. $Q_\theta$ is the noise intensity, $\tau_\theta$  the self-correlation time.

\begin{figure}
\centering
\subfigure{
\includegraphics[height=5.5cm,width=7cm]{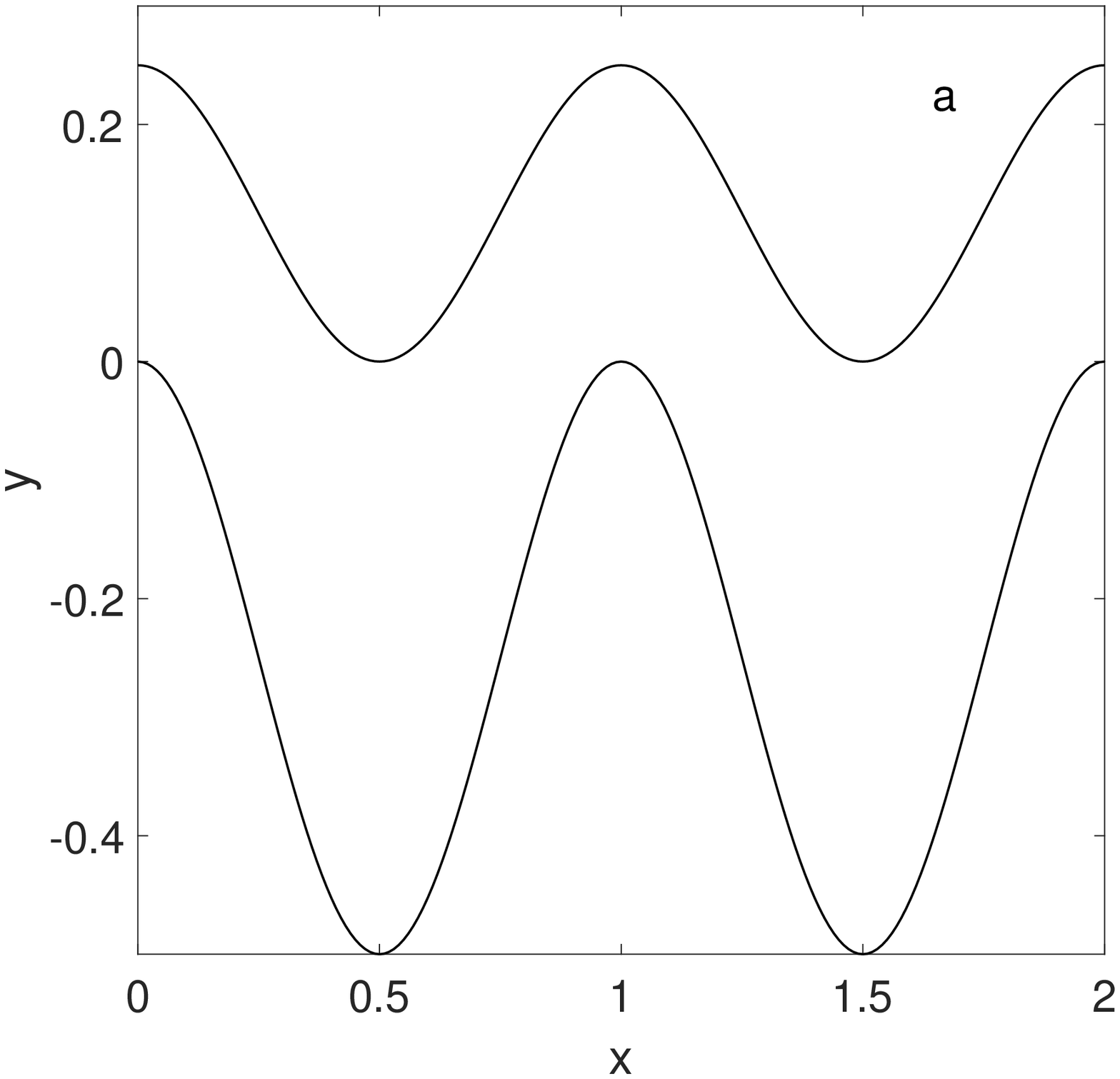}
}
\subfigure{
\includegraphics[height=5.5cm,width=7cm]{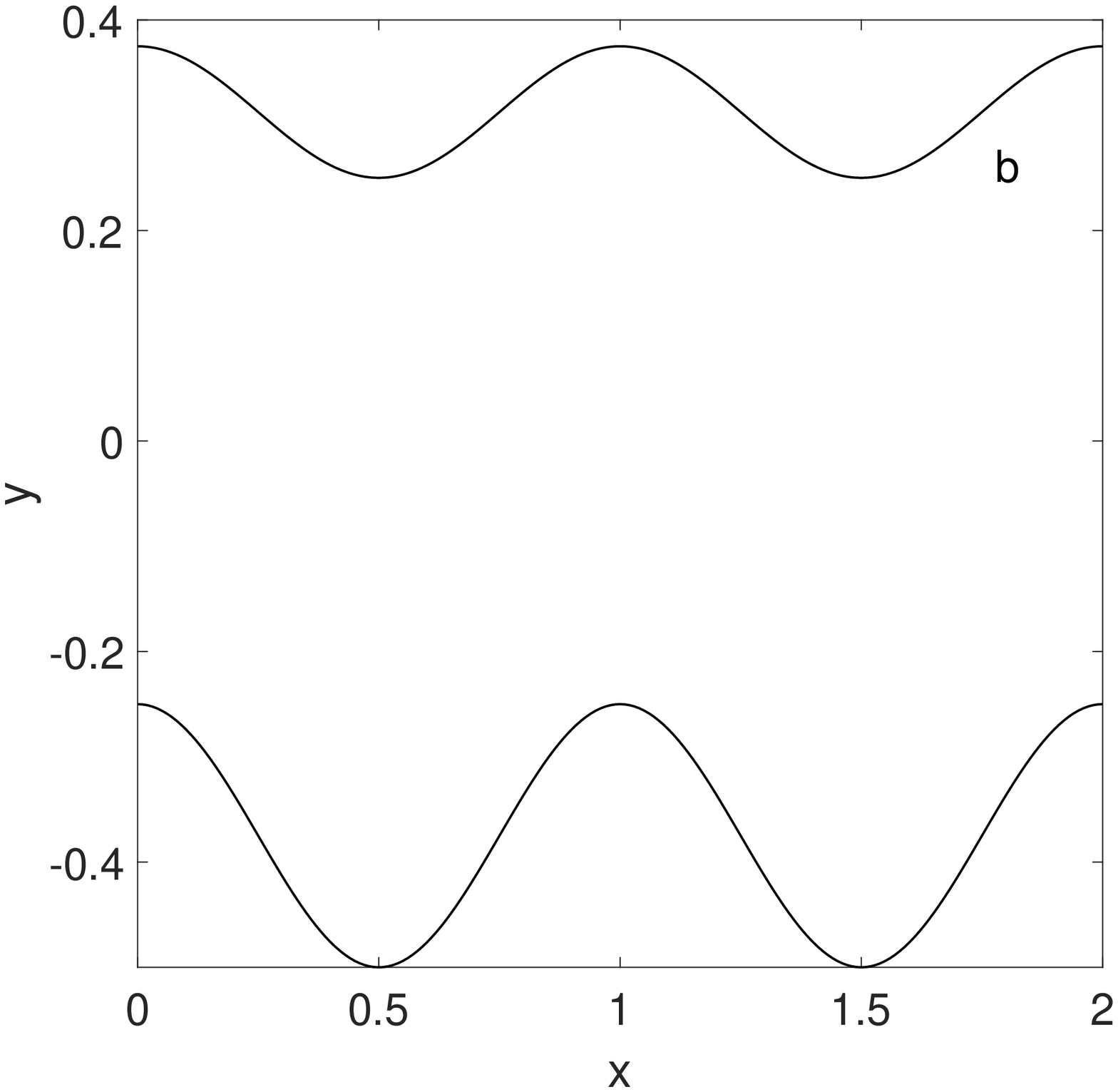}
}
\subfigure{
\includegraphics[height=5.5cm,width=7cm]{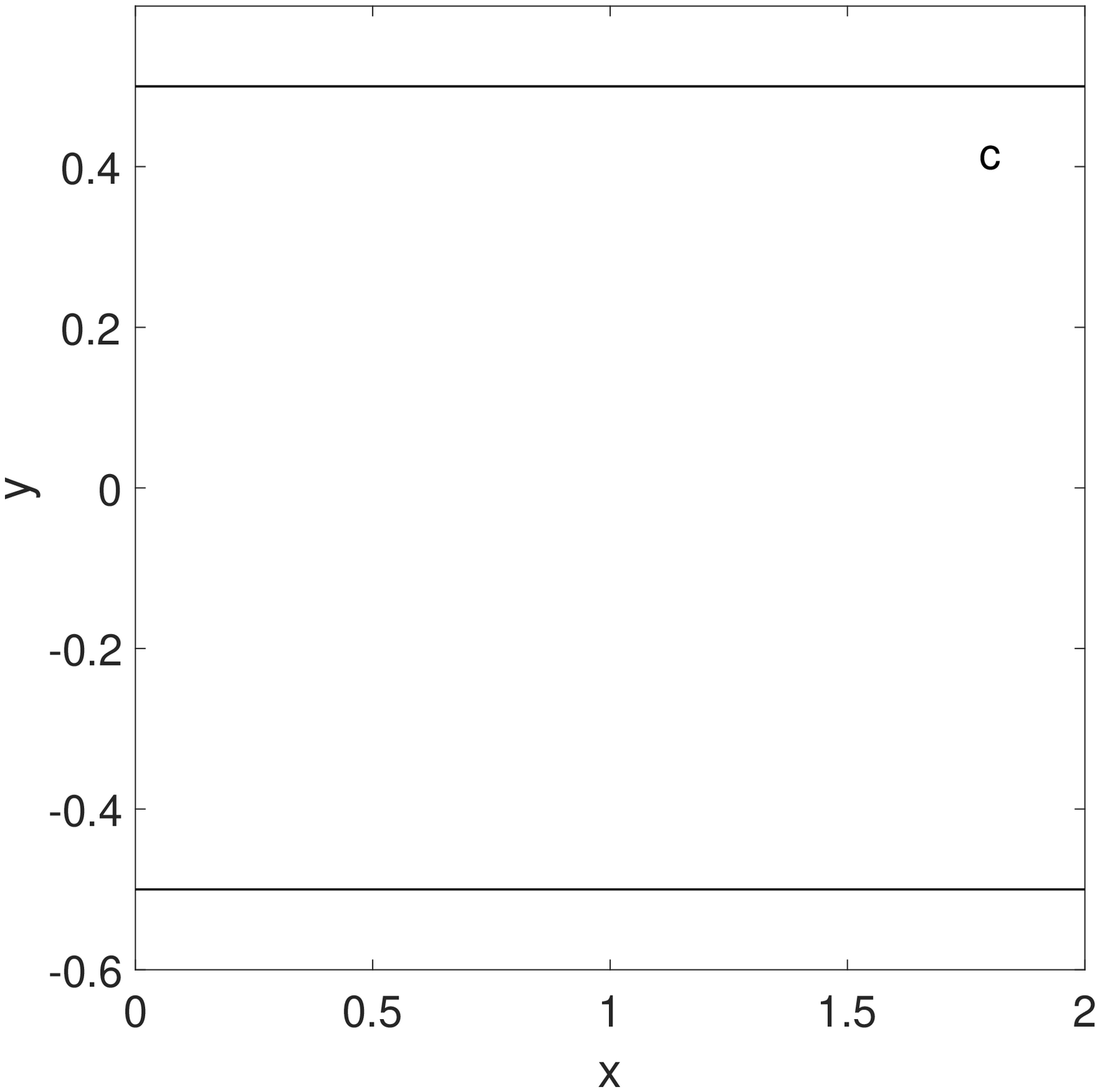}
}
\subfigure{
\includegraphics[height=5.5cm,width=7cm]{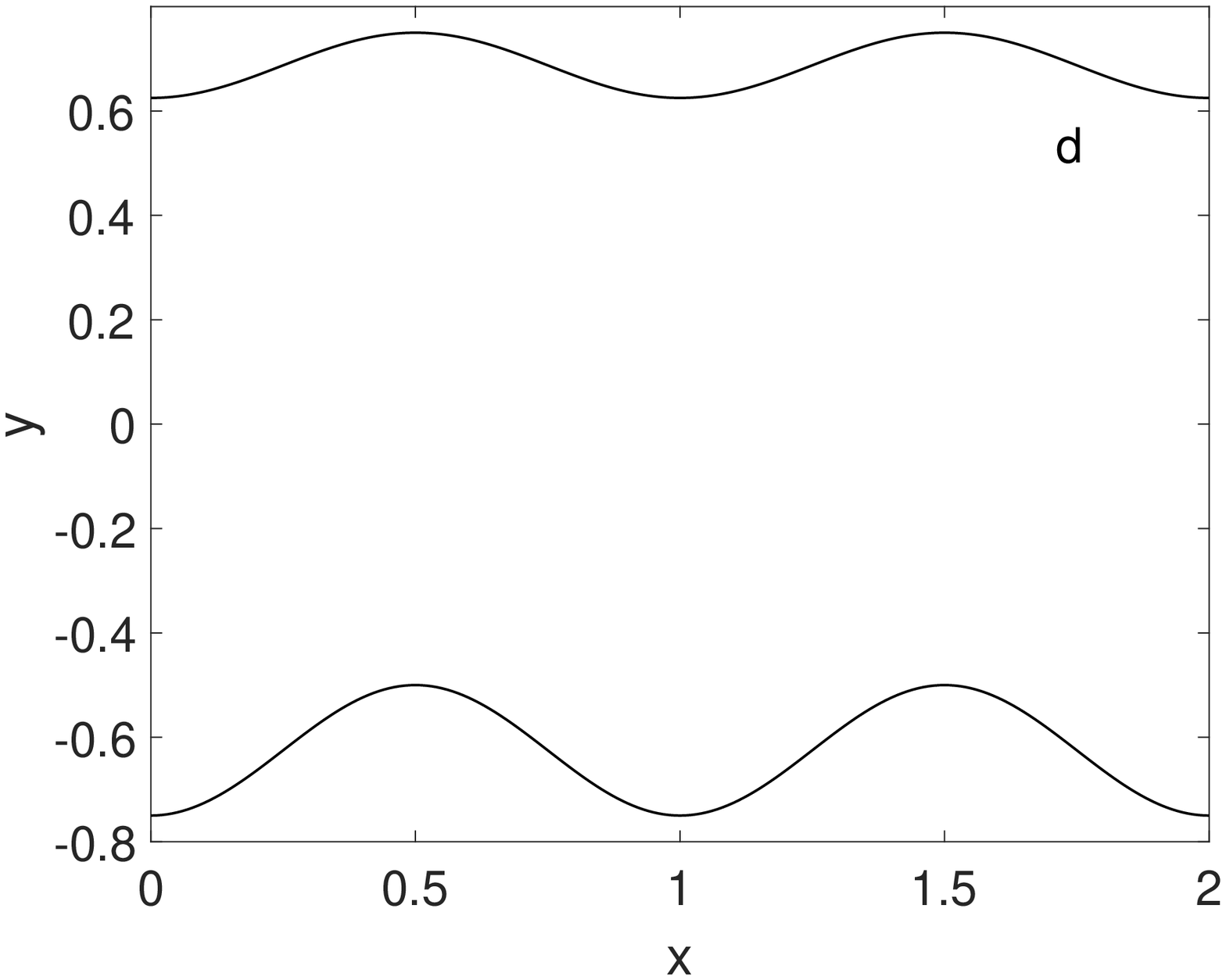}
}
\subfigure{
\includegraphics[height=5.5cm,width=7cm]{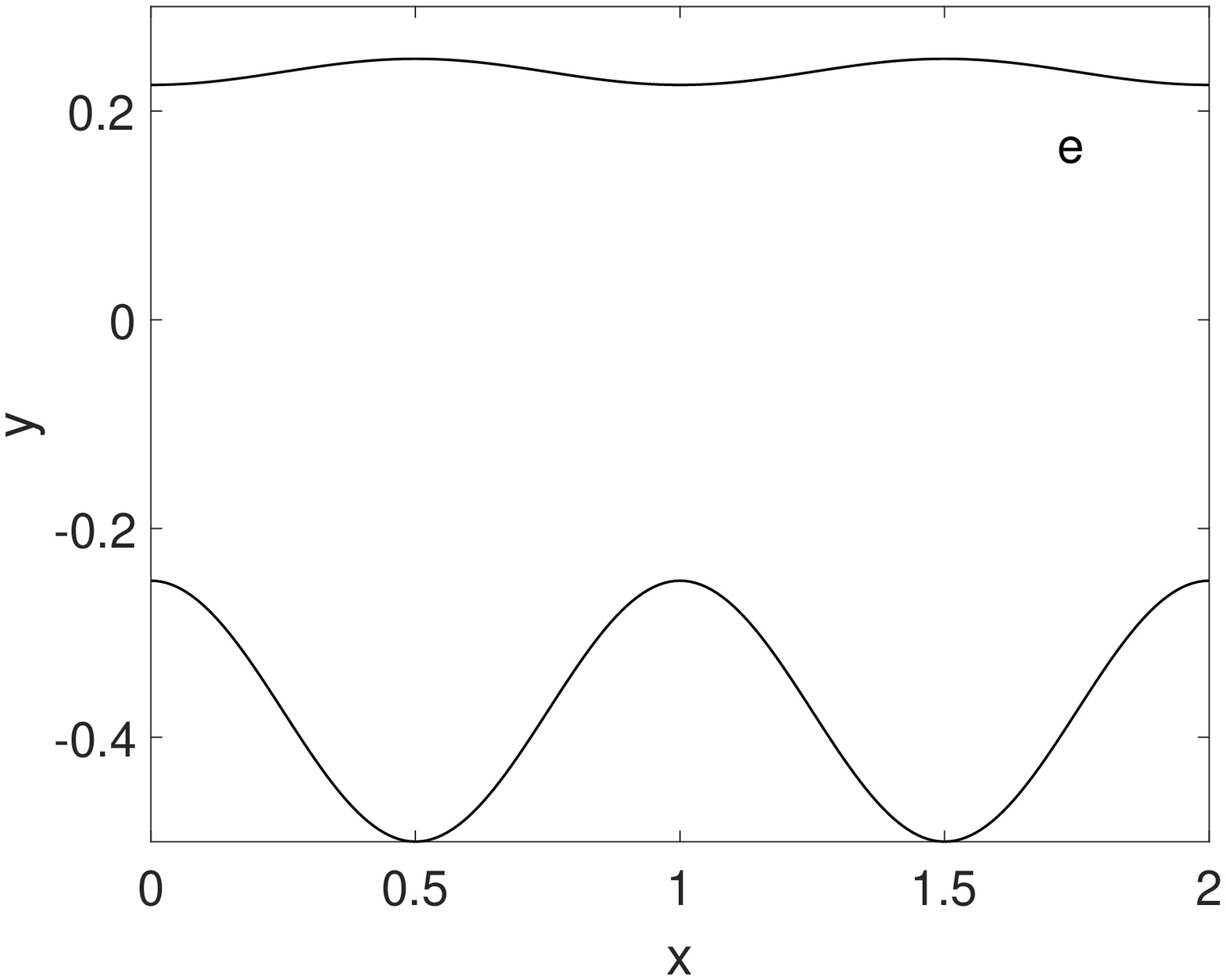}
}
\subfigure{
\includegraphics[height=5.5cm,width=7cm]{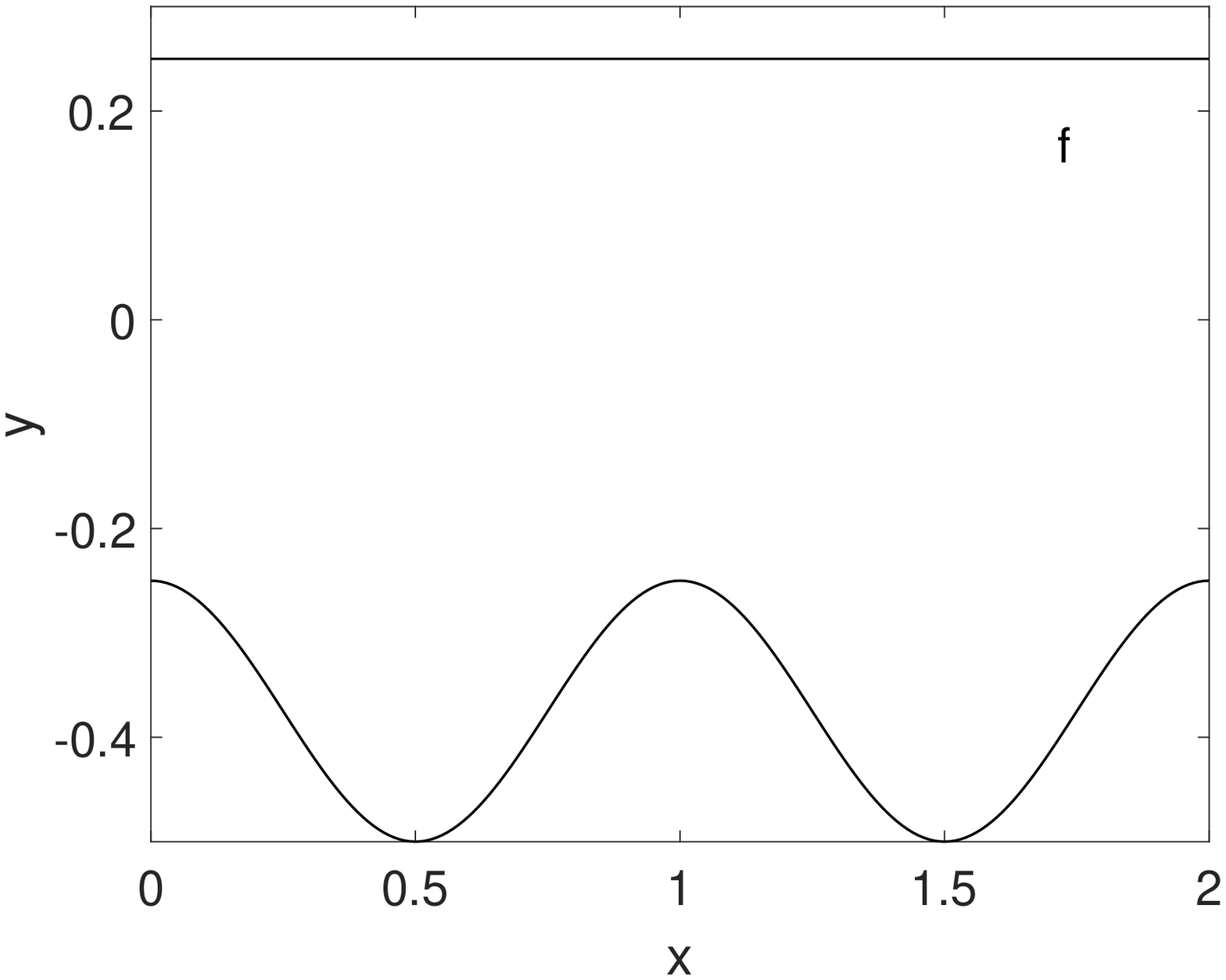}
}
\subfigure{
\includegraphics[height=5.5cm,width=7cm]{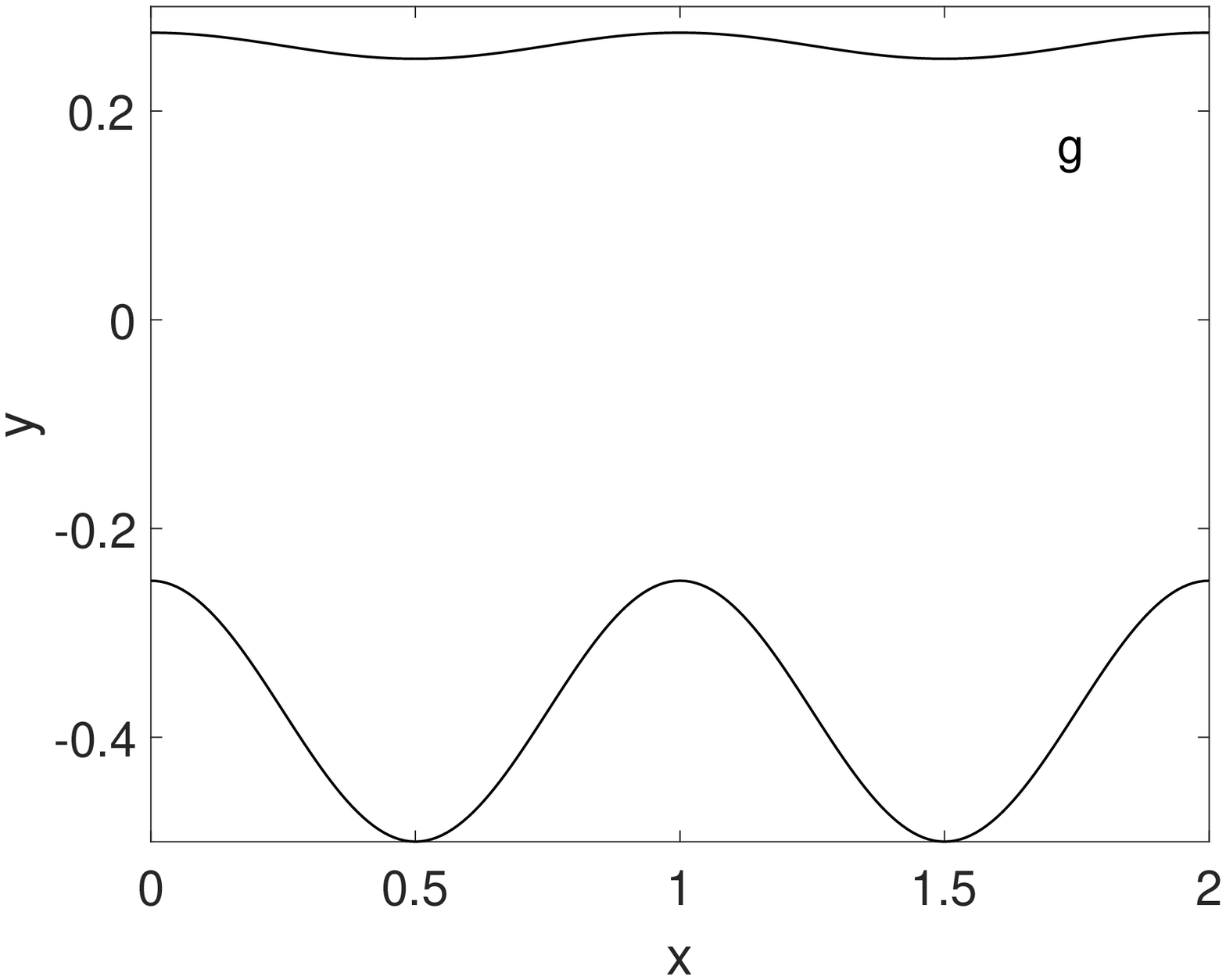}
}
\subfigure{
\includegraphics[height=5.5cm,width=7cm]{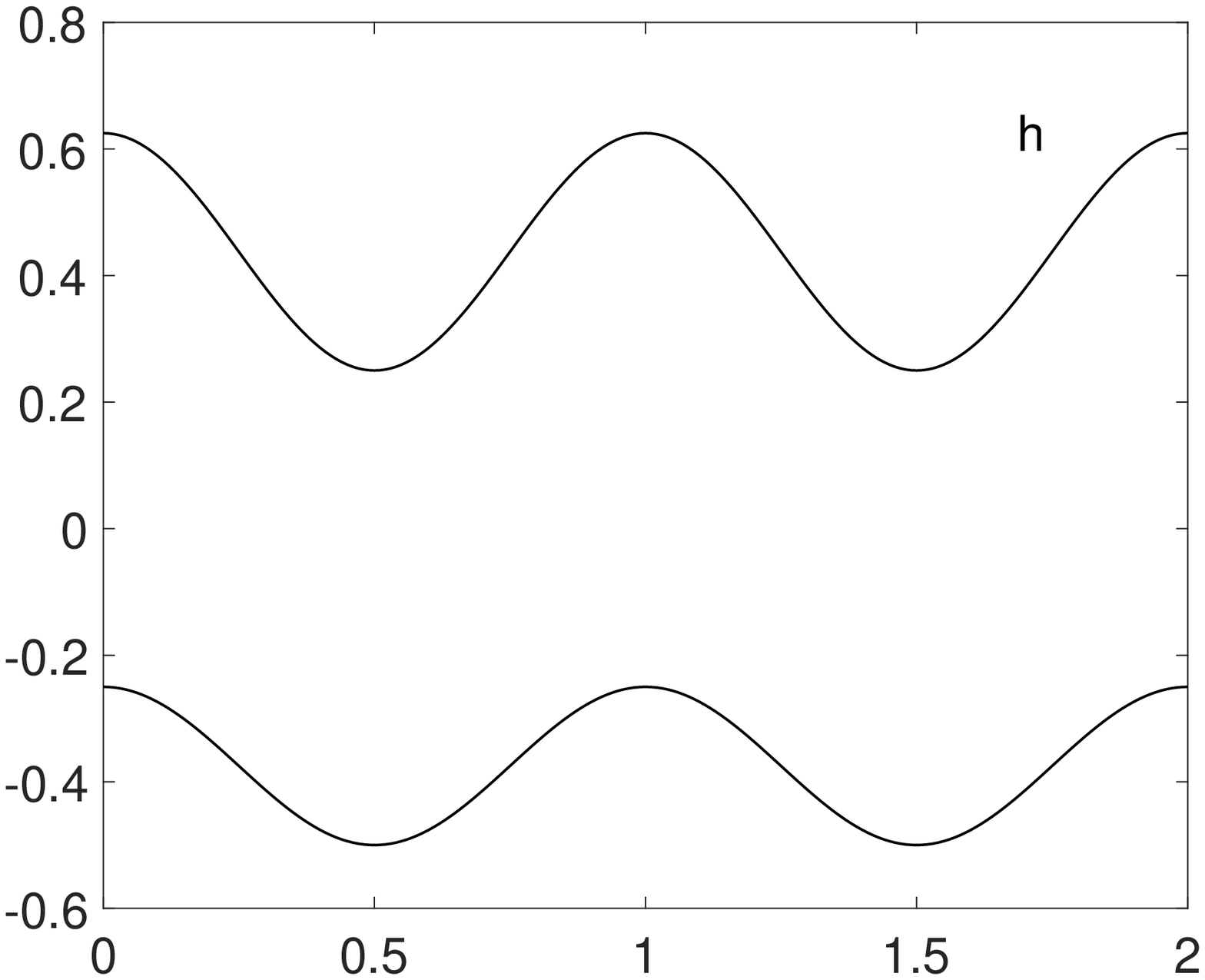}
}
\caption{Schematic of the corrugated channel with different asymmetries:(a)$\epsilon=0.5$, $\Delta=0.0$, $\phi=\pi$; (b)$\epsilon=0.5$, $\Delta=0.5$, $\phi=\pi$; (c) $\epsilon=0.5$, $\Delta=1.0$, $\phi=\pi$; (d)$\epsilon=0.5$, $\Delta=1.5$, $\phi=\pi$; (e)$\epsilon=-0.1$, $\Delta=0.5$, $\phi=\pi$; (f)$\epsilon=0.0$, $\Delta=0.5$, $\phi=\pi$; (g)$\epsilon=0.1$, $\Delta=0.5$, $\phi=\pi$; (h)$\epsilon=1.5$, $\Delta=0.5$, $\phi=\pi$.}
\label{Channel}
\end{figure}
The confined corrugated channel is a periodic function in space along the $x$ direction(depicted in Fig.\ref{Channel}). The walls of the cavity have been modelled by the following function

\begin{eqnarray}
W_+(x)=\frac{1}{2}\{\Delta +\epsilon(y_L-\Delta)\sin^2(\frac{\pi }{x_L}x+\frac{\phi}{2})\}\\
W_-(x)=-\frac{1}{2}\{\Delta +(y_L-\Delta)\sin^2(\frac{\pi }{x_L}x)\}
\end{eqnarray}

The upper and lower boundary functions are $W_+(x)$ and $W_-(x)$, respectively. $x_L$ is the compartment length, $y_L$ the channel width, and $\Delta$ the bottleneck size. There are two geometrical parameters introduced in $W_+(x)$ for varying the upside-down asymmetric degree, namely $\epsilon$ and $\phi$. $\epsilon$ is defined as real number for tuning the amplitude of the upper wall compared to the lower wall, and $\phi$ is for tuning the shift of the upper wall from corresponding position of lower wall.

A central practical question in the theory of Brownian motors is the over all long time behavior of the particle, and the key quantities of particle transport is the particle velocity $\langle V\rangle$. Because particles along the $y$ direction are confined, we only calculate the $x$ direction average velocity
\begin{equation}
\langle V_{\theta_0}\rangle=\lim_{t\to\infty}\frac{\langle{x(t)-x(t_0)}\rangle}{t-t_0}
\end{equation}
$x(t_0)$ is the position of particles at time $t_0$.
$\theta_0$ is initial angle of the trajectory. The full average velocity after another average over all $\theta_0$ is
\begin{equation}
\langle V\rangle=\frac{1}{2\pi}\int^{2\pi}_0 \langle V_{\theta_0}\rangle d\theta_0
\end{equation}

\section{\label{label3}Results and discussion}

In order to give a simple and clear analysis of the system. Eqs.(\ref{Ext}), (\ref{Eyt})and (\ref{Ethetat}) are integrated using the Euler algorithm. In our simulations, the integration step time $\Delta t=10^{-4}$ and the total integration time was more than $5\times 10^5$ and the transient effects were estimated and subtracted. The stochastic averages reported above were obtained as ensemble averages over $2\times 10^4$ trajectories with random initial conditions.

\begin{figure}
\centering
\subfigure{
\includegraphics[height=8cm,width=10cm]{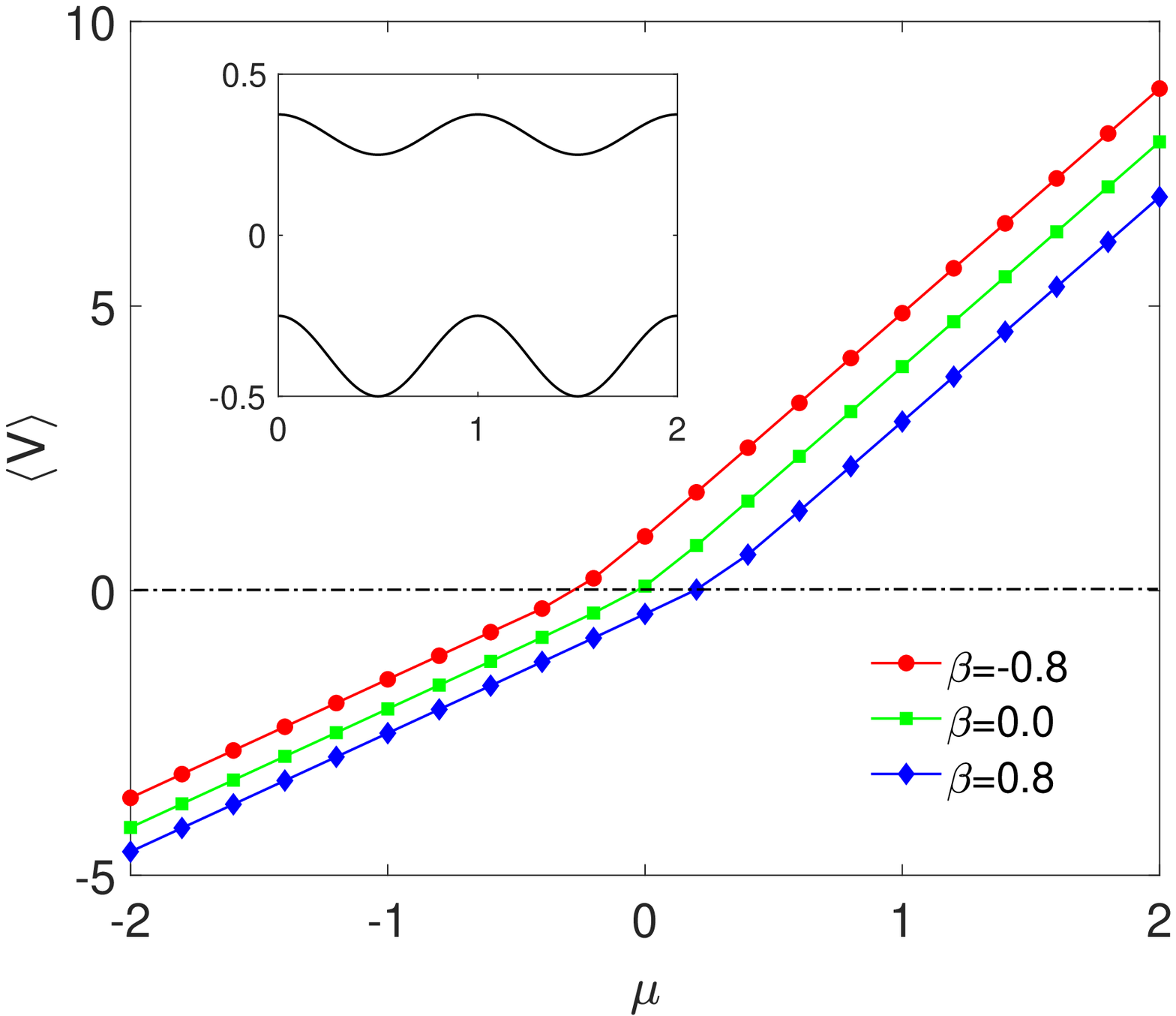}
}
\caption{The average velocity $\langle V\rangle$ as a function of mean parameter $\mu$ with different asymmetry parameter $\beta$. The other parameters are $\epsilon=0.5$, $\Delta=0.5$, $\phi=\pi$, $\alpha=1.2$, $\sigma=0.5$, $v_0=0.1$, $\omega=0.2$, $x_L=y_L=1.0$, $Q_\theta=0.2$, $\tau_{\theta}=1.0$.}
\label{VMu}
\end{figure}

We known that the mean parameter $\mu$ shifts the distribution to the left($\mu<0$) or right($\mu>0$). The average velocity $\langle V\rangle$ as a function of mean parameter $\mu$ with different asymmetry parameter $\beta$ is reported in Fig.\ref{VMu}. In this figure, we find $\langle V\rangle$ increases with increasing $\mu$. The moving direction changes from against $x$ axis to along $x$ axis with increasing $\mu$, so the transport reverse phenomenon appears with increasing mean parameter $\mu$. The transport reverse point coordinates $\mu_{trp}<0$ when $\beta=-0.8$, and $\mu_{trp}=0$ when $\beta=0$, and $\mu_{trp}>0$ when $\beta=0.8$. When the asymmetry parameter $\beta=0$, left distribution induces $-x$ directional transport, and right distribution induces $x$ directional transport. When $\beta<0$($\beta=-0.8$), the turning point moves to left because the distribution is skewed to left. When $\beta>0$($\beta=0.8$), the turning point moves to right as the distribution is skewed to right. In this figure, we also find the smaller $\beta$, the larger $\langle V\rangle$ is.

\begin{figure}
\centering
\subfigure{
\includegraphics[height=8cm,width=10cm]{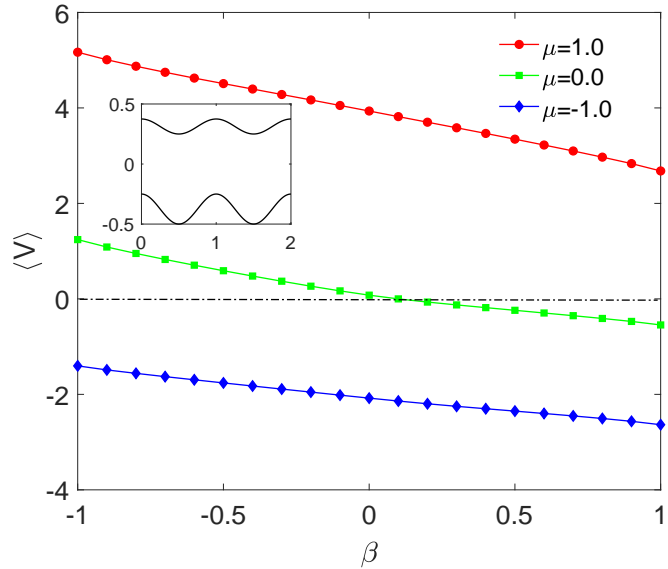}
}
\caption{The average velocity $\langle V\rangle$ as a function of asymmetry parameter $\beta$ with different mean parameter $\mu$. The other parameters are $\epsilon=0.5$, $\Delta=0.5$, $\phi=\pi$, $\alpha=1.2$, $\sigma=0.5$, $v_0=0.1$, $\omega=0.2$, $x_L=y_L=1.0$, $Q_\theta=0.2$, $\tau_{\theta}=1.0$.}
\label{VBeta}
\end{figure}

The asymmetry parameter $\beta$($-1\leq\beta\leq1$) determines the skewness of the distribution. The average velocity $\langle V\rangle$ as a function of $\beta$ with different $\mu$ is reported in Fig.\ref{VBeta}. When $\mu=1.0$(right distribution), $\langle V\rangle>0$ and $\langle V\rangle$ decreases with increasing $\beta$, so right skewed distribution will inhibit the particle transport in $x$ direction. When $\mu=0.0$, with increasing $\beta$, the moving direction changes from along $x$ axis to against $x$ axis, and the turning point is $\beta=0$. When $\mu=-1.0$(left distribution), $\langle V\rangle<0$ and the moving speed $|\langle V\rangle|$ increases with increasing $\beta$, so large $\beta$ is good for particle directional transport in $-x$ direction. We know left(right) distribution induces $-x$($+x$) directional transport(in figure \ref{VMu}), but left(right) skewed distribution induces $+x$($-x$) directional transport, so the moving direction is determined by the joint effects of $\mu$ and $\beta$.

\begin{figure}
\centering
\subfigure{
\includegraphics[height=6cm,width=7cm]{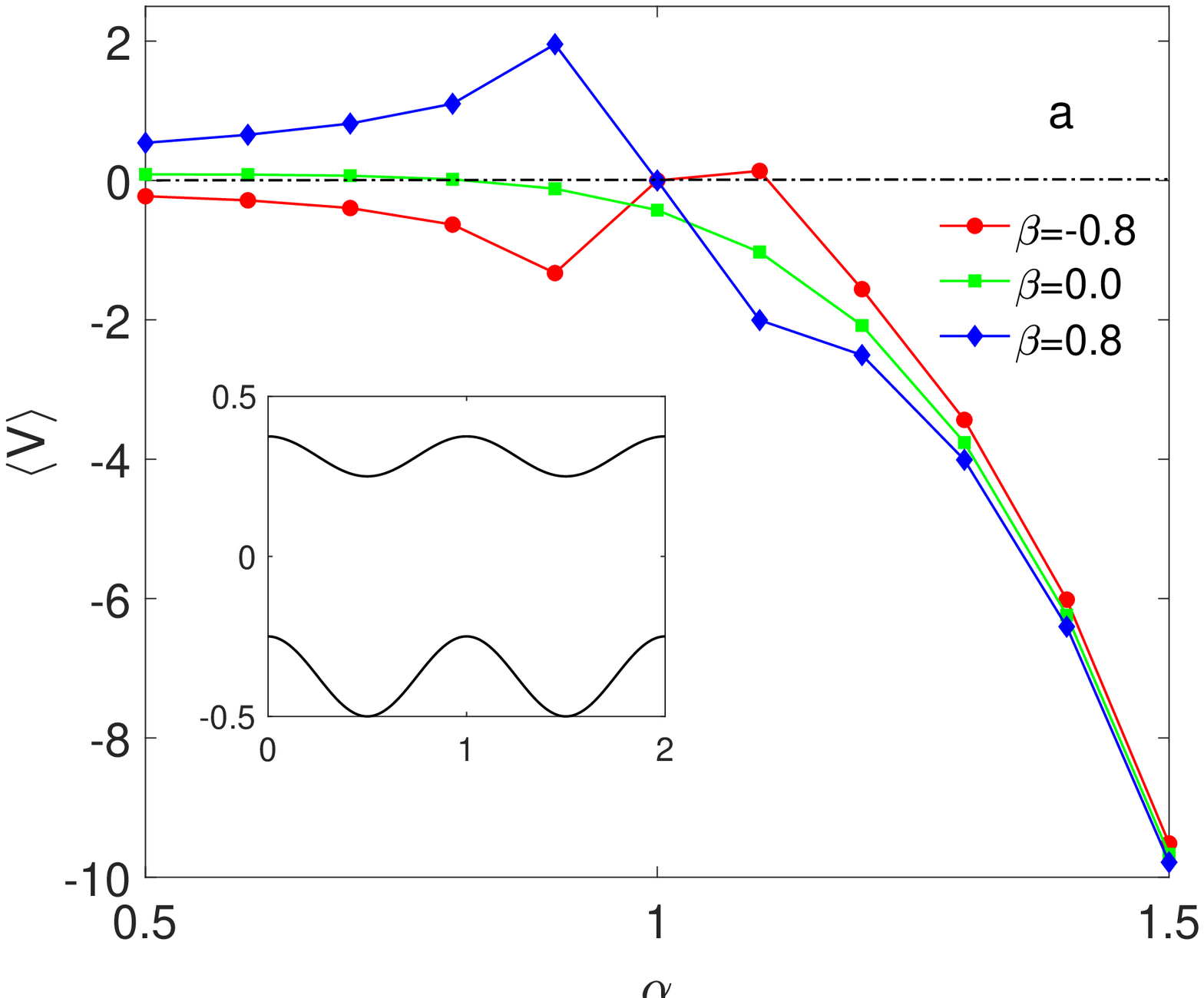}
}
\subfigure{
\includegraphics[height=6cm,width=7cm]{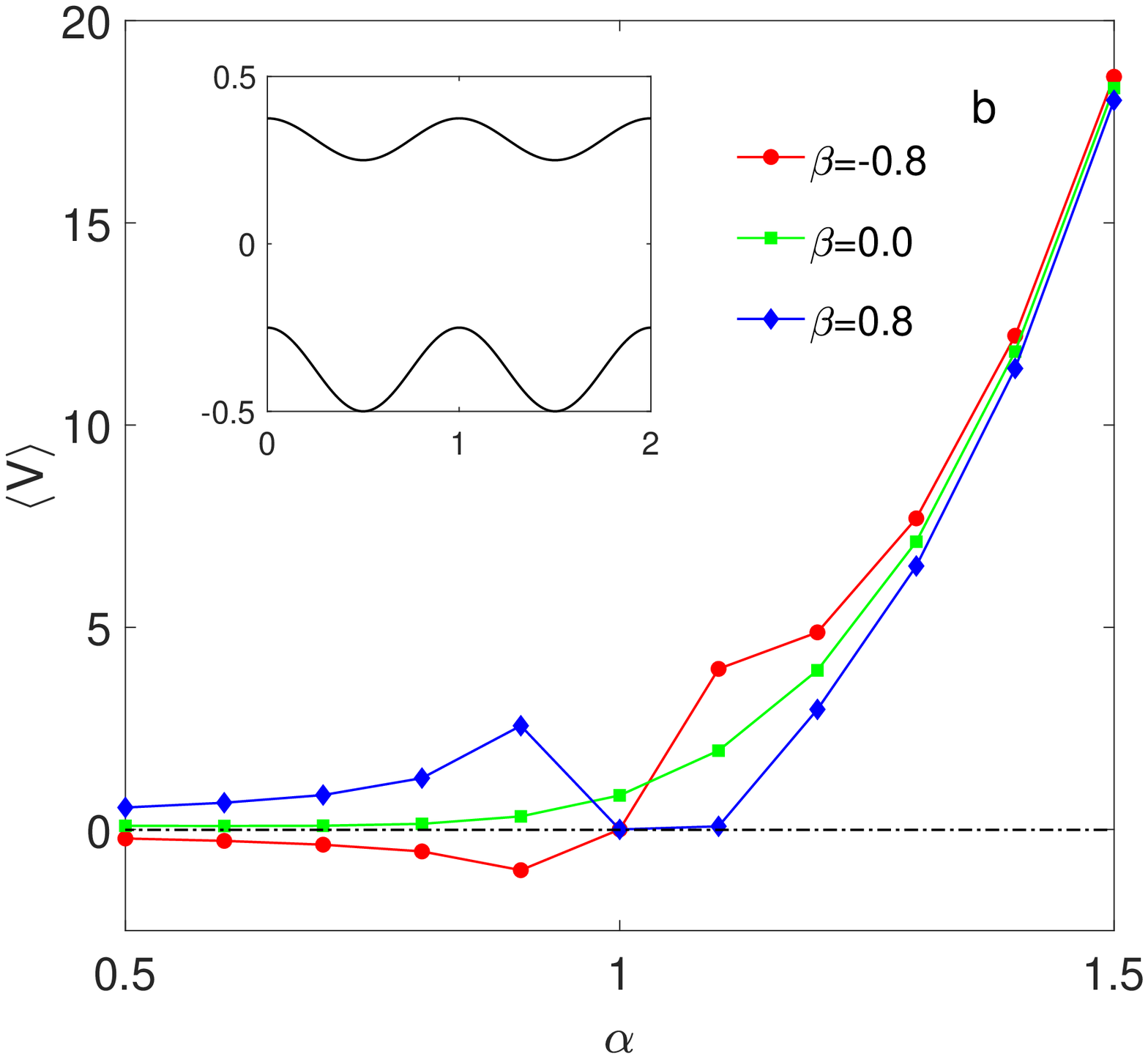}
}
\caption{The average velocity $\langle V\rangle$ as a function of stability index $\alpha$ with different asymmetry parameter $\beta$. The other parameters are $\epsilon=0.5$, $\Delta=0.5$, $\phi=\pi$, $\sigma=0.5$, $v_0=0.1$, $\omega=0.2$, $Q_\theta=0.2$, $\tau_{\theta}=1.0$:(a)$\mu=-1.0$, (b)$\mu=1.0$.}
\label{VAlpha}
\end{figure}

The average velocity $\langle V\rangle$ as a function of stability index $\alpha$ with different asymmetry $\beta$ is reported in Fig.\ref{VAlpha}. It is found that $\langle V\rangle$ shows complex behavior with increasing $\alpha$. In Fig.\ref{VAlpha}(a)($\mu=-1$), when $\beta=-0.8$, $\langle V\rangle$ decreases with increasing $\alpha$ and reaches a minimum when $\alpha=0.9$, then $\langle V\rangle$ increases with increasing $\alpha$ and reaches a maximum when $\alpha=1.1$, and then $\langle V\rangle$ decreases with increasing $\alpha$ when $\alpha>1.1$. In Fig.\ref{VAlpha}(a), when $\beta=0.0$, $\langle V\rangle$ decreases monotonically with increasing $\alpha$, so large $\alpha$ is good for particle directional transport in $-x$ direction. In Fig.\ref{VAlpha}(a), when $\beta=0.8$, $\langle V\rangle$ has a maximum with increasing $\alpha$ and the transport reverse phenomenon appears with increasing $\alpha$. In Fig.\ref{VAlpha}(b)($\mu=1$), when $\beta=-0.8$, there exists a minimum with increasing $\alpha$ and the transport reverse phenomenon appears with increasing $\alpha$. In Fig.\ref{VAlpha}(b), when $\beta=0.0$, $\langle V\rangle>0$ and $\langle V\rangle$ increases monotonically with increasing $\alpha$, so large stability index is good for particle directional transport in $x$ direction. In Fig.\ref{VAlpha}(b), when $\beta=0.8$, there exit a maximum and a minimum with increasing $\alpha$. In this figure, we find the $\langle V\rangle-\alpha$ lines will coincide with each other when $\alpha\geq1.3$, the reasons for this is the larger stability, the stronger impact of the noise.

\begin{figure}
\centering
\subfigure{
\includegraphics[height=6cm,width=7cm]{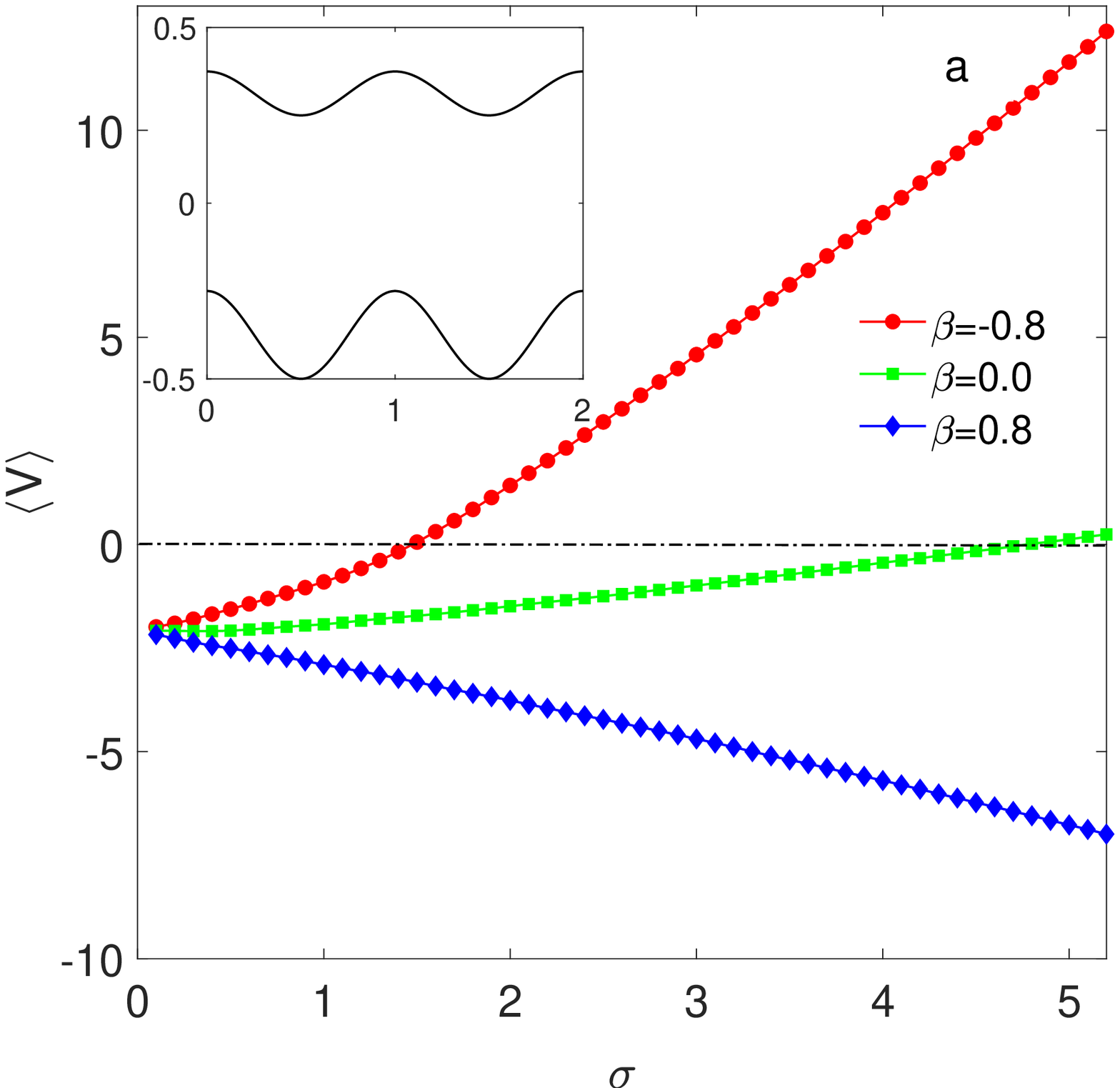}
}
\subfigure{
\includegraphics[height=6cm,width=7cm]{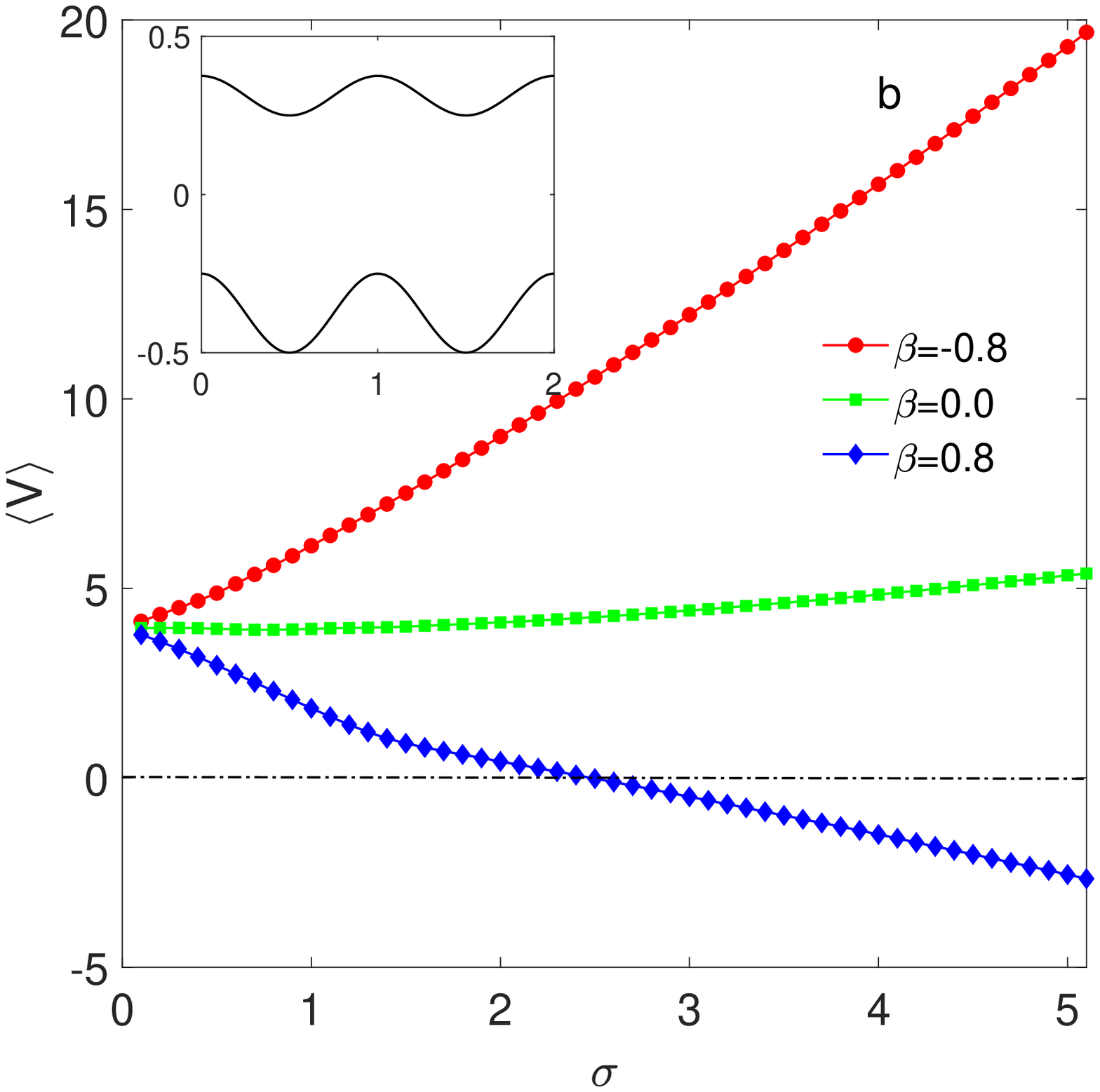}
}
\caption{The average velocity $\langle V\rangle$ as a function of scale parameter $\sigma$ with different asymmetry parameter $\beta$. The other parameters are $\epsilon=0.5$, $\Delta=0.5$, $\phi=\pi$, $\alpha=1.2$, $v_0=0.1$, $\omega=0.2$, $x_L=y_L=1.0$, $Q_\theta=0.2$, $\tau_{\theta}=1.0$:(a)$\mu=-1.0$, (b)$\mu=1.0$.}
\label{VSigma}
\end{figure}

The average velocity $\langle V\rangle$ as a function of scale parameter $\sigma$ with different $\beta$ is reported in Fig.\ref{VSigma}. In Fig.\ref{VSigma}(a)($\mu=-1.0$), when $\beta=-0.8$ and $\beta=0.0$, we find $\langle V\rangle$ increases with increasing $\sigma$ and the transport reverse phenomenon appears with increasing $\sigma$, so small value of $\sigma$ will help to particle directional transport in $-x$ direction, but large $\sigma$ will help to particle transport in $x$ direction. In Fig.\ref{VSigma}(a), when $\beta=0.8$, $\langle V\rangle<0$ and $\langle V\rangle$ decreases with increasing $\sigma$, so large $\sigma$ will help to particle directional transport in $-x$ direction. In Fig.\ref{VSigma}(b)($\mu=1.0$), when $\beta=-0.8$ and $\beta=0.0$, the particle moves in $x$ direction and large $\sigma$ helps to particle direction transport($\langle V\rangle>0$ and $\langle V\rangle$ increases with increasing $\sigma$). When $\beta=0.8$, $\langle V\rangle$ decreases with increasing $\sigma$ and the transport reverse phenomenon appears.

\begin{figure}
\centering
\subfigure{
\includegraphics[height=6cm,width=7cm]{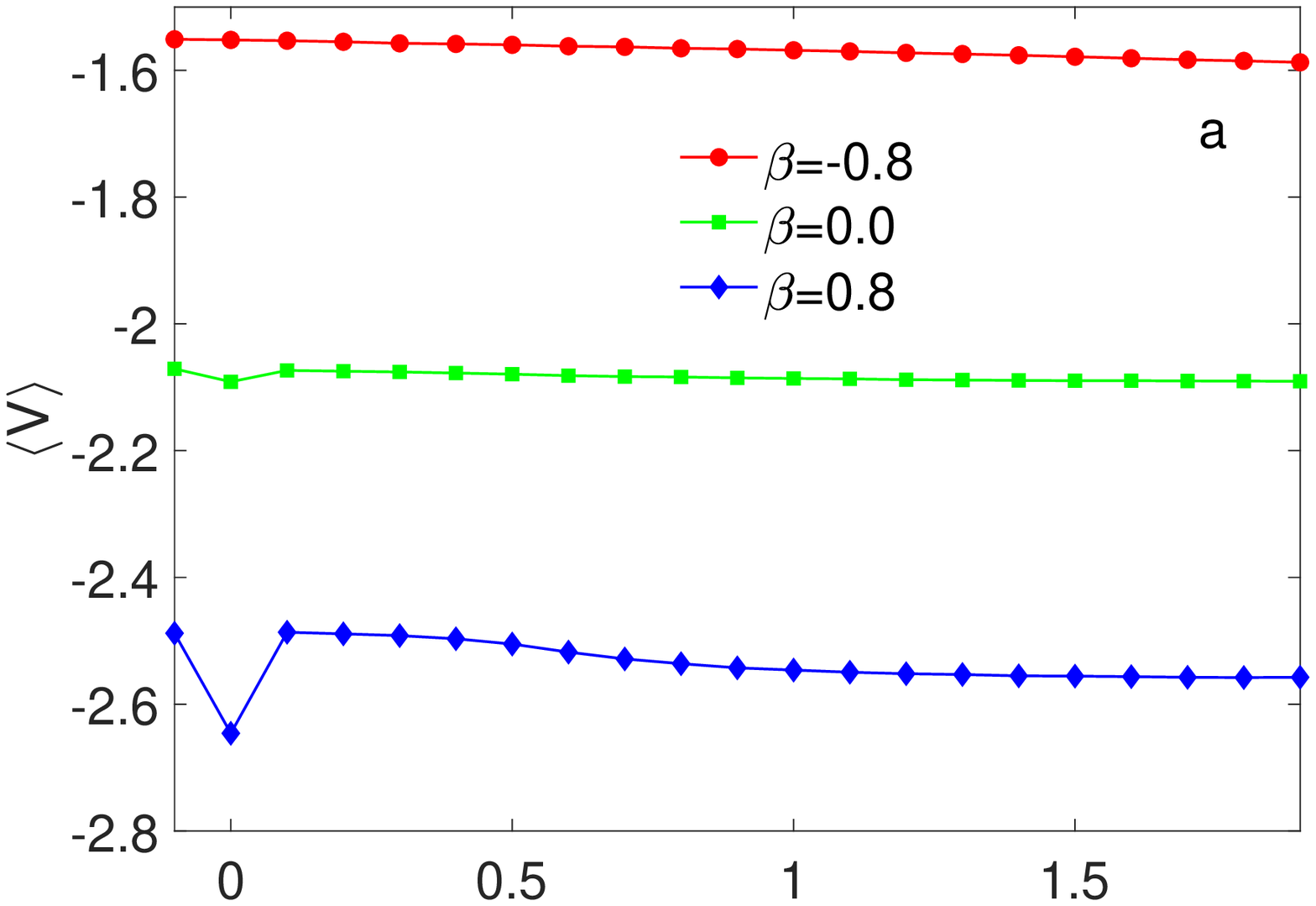}
}
\subfigure{
\includegraphics[height=6cm,width=7cm]{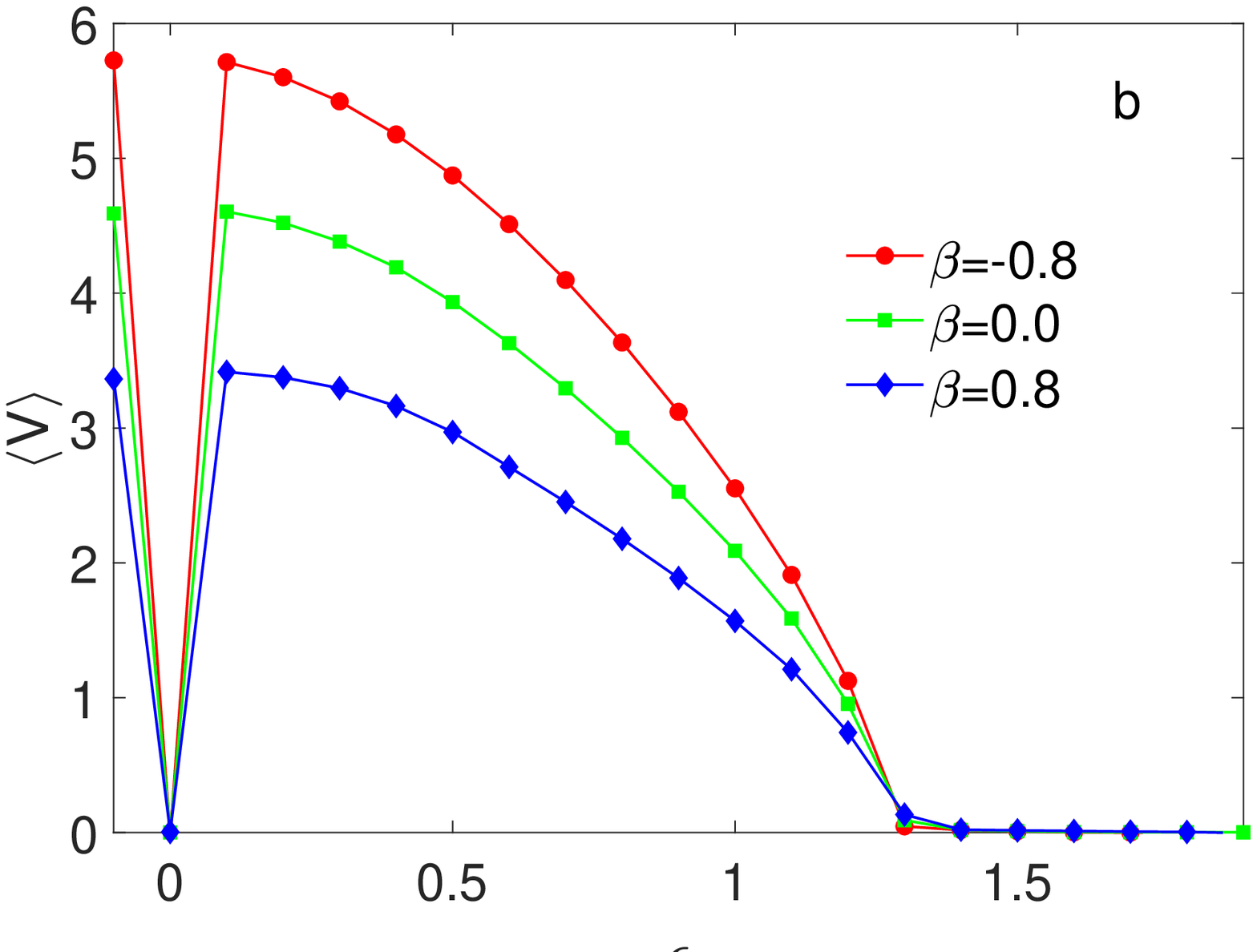}
}
\caption{The average velocity $\langle V\rangle$ as a function of $\epsilon$ with different asymmetry parameter $\beta$. The other parameters are $\Delta=0.5$, $\phi=\pi$, $\alpha=1.2$, $\sigma=0.5$, $v_0=0.1$, $\omega=0.2$, $x_L=y_L=1.0$, $Q_\theta=0.2$, $\tau_{\theta}=1.0$:(a)$\mu=-1.0$, (b)$\mu=1.0$.}
\label{Vepsilon}
\end{figure}

Fig.\ref{Vepsilon} shows the average velocity $\langle V\rangle$ as a function of $\epsilon$ with different $\beta$. In the case of $\mu=-1.0$(Fig.\ref{Vepsilon}(a)), $\langle V\rangle$ decreases slowly with increasing $\epsilon$ when $\beta=-0.8$. When $\beta=0.0$, there exists a unconspicuous minimum($\langle V\rangle_{min}\approx-2.09$ when $\epsilon=0$) with increasing $\epsilon$. When $\beta=0.8$, there exists an obvious minimum($\langle V\rangle_{min}\approx-2.65$ when $\epsilon=0$) with increasing $\epsilon$ and $\langle V\rangle$ decreases with increasing $\epsilon$ when $\epsilon>0.1$. In the case of $\mu=1.0$(Fig.\ref{Vepsilon}(b)), we find $\langle V\rangle\geq0$ and $\langle V\rangle$ shows complex appearance with increasing $\epsilon$. The $\langle V\rangle-\epsilon$ curve has the same shape for different $\beta$($\beta=-0.8$, $\beta=0.0$ and $\beta=0.8$). $\langle V\rangle$ has a large value when $\epsilon=-0.1$, and decreases quickly to zero when $\epsilon=0.0$, and then increases quickly to a large value when $\epsilon=0.1$, and decrease to small value(about zero) when $\epsilon>1.3$.

\begin{figure}
\centering
\subfigure{
\includegraphics[height=6cm,width=7cm]{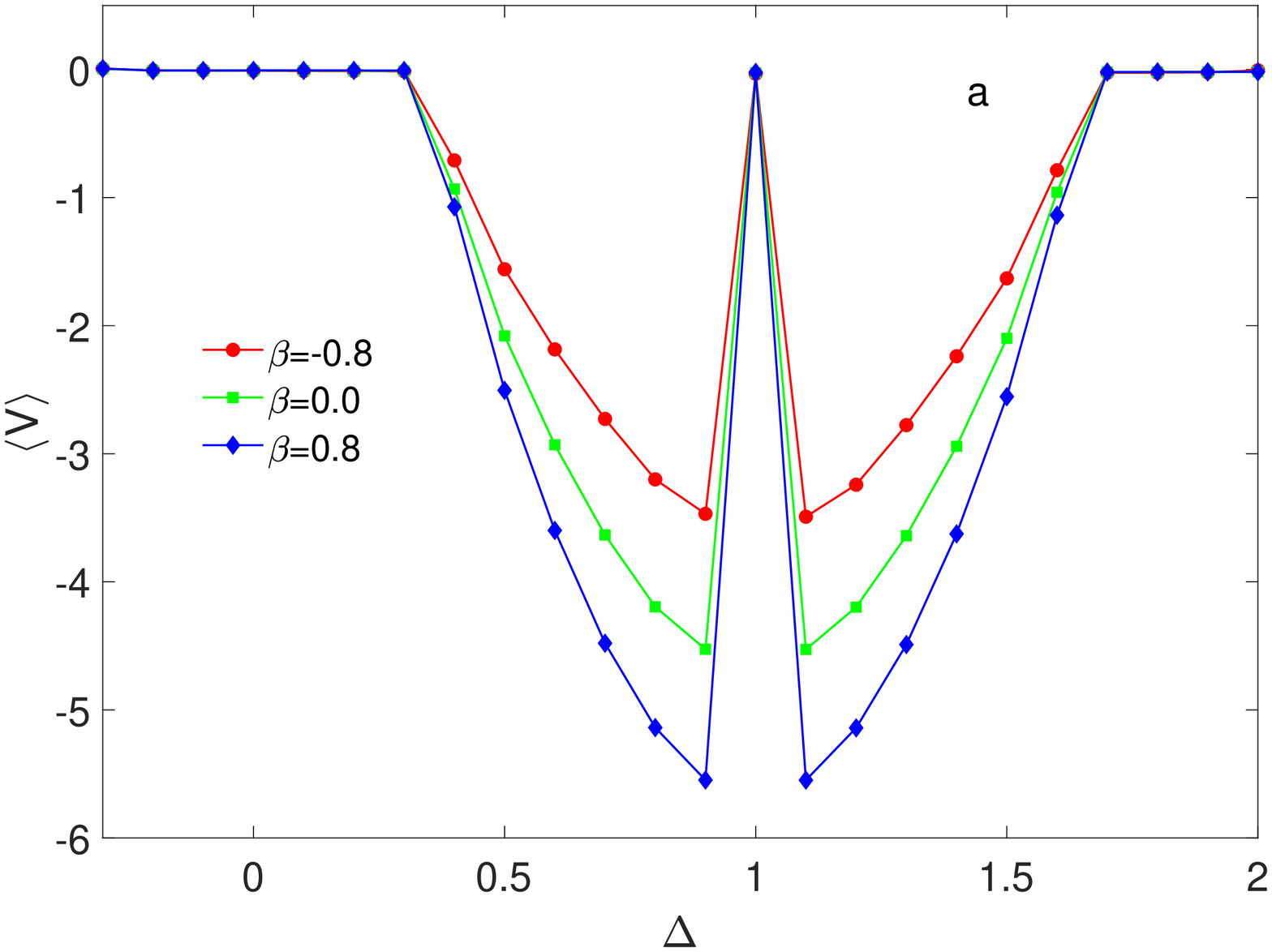}
}
\subfigure{
\includegraphics[height=6cm,width=7cm]{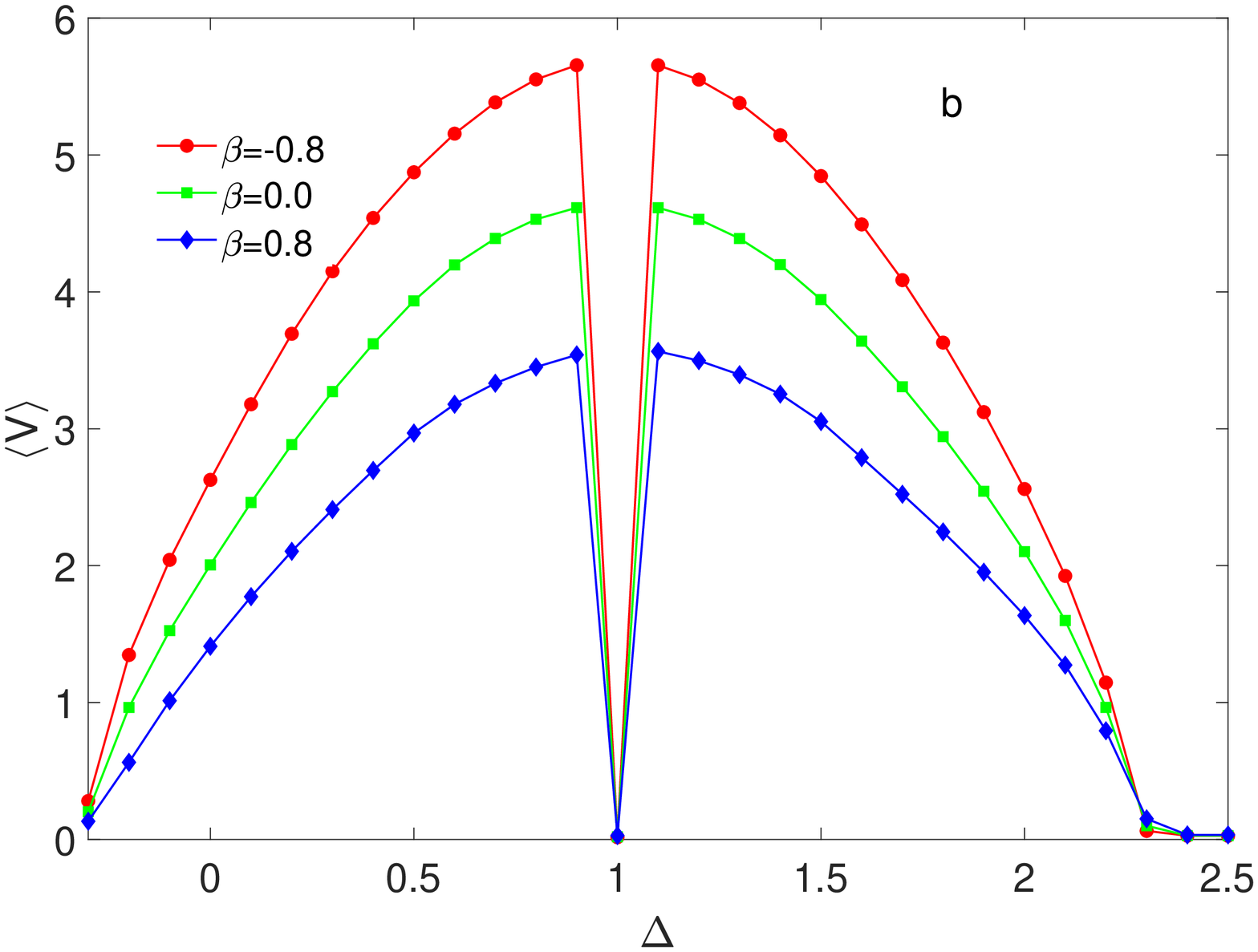}
}
\caption{The average velocity $\langle V\rangle$ as a function of bottleneck size $\Delta$ with different asymmetry parameter $\beta$. The other parameters are $\epsilon=0.5$, $\phi=\pi$, $\alpha=1.2$, $\sigma=0.5$, $v_0=0.1$, $\omega=0.2$, $x_L=y_L=1.0$, $Q_\theta=0.2$, $\tau_{\theta}=1.0$:(a)$\mu=-1.0$, (b)$\mu=1.0$.}
\label{VDelta}
\end{figure}

Fig.\ref{VDelta} shows $\langle V\rangle$ as a function of $\Delta$ with different $\beta$. When $\mu=-1.0$(Fig.\ref{VDelta}(a)), $\langle V\rangle\leq0$. When $\mu=1.0$(Fig.\ref{VDelta}(b)), $\langle V\rangle\geq0$. In Fig.\ref{VDelta}(a), $\langle V\rangle\approx0$ when $\Delta\leq0.3$, and $\langle V\rangle$ decreases to a minimum when $\Delta=0.9$, and increases to zero when $\Delta=1.0$(whenever $\beta=0.8$, $\beta=0.0$ and $\beta=0.8$), and decreases to another minimum when $\Delta=1.1$, and then increases to zero again when $\Delta=1.6$(whenever $\beta=0.8$, $\beta=0.0$ and $\beta=0.8$). Generally, the $\langle V\rangle-\Delta$ curve forms the shape of word "W" when $\mu=-1.0$. In the case of $\mu=1.0$(Fig.\ref{VDelta}(b)), $\langle V\rangle-\Delta$ curve forms the shape of word "M".

\begin{figure}
\centering
\subfigure{
\includegraphics[height=6cm,width=7cm]{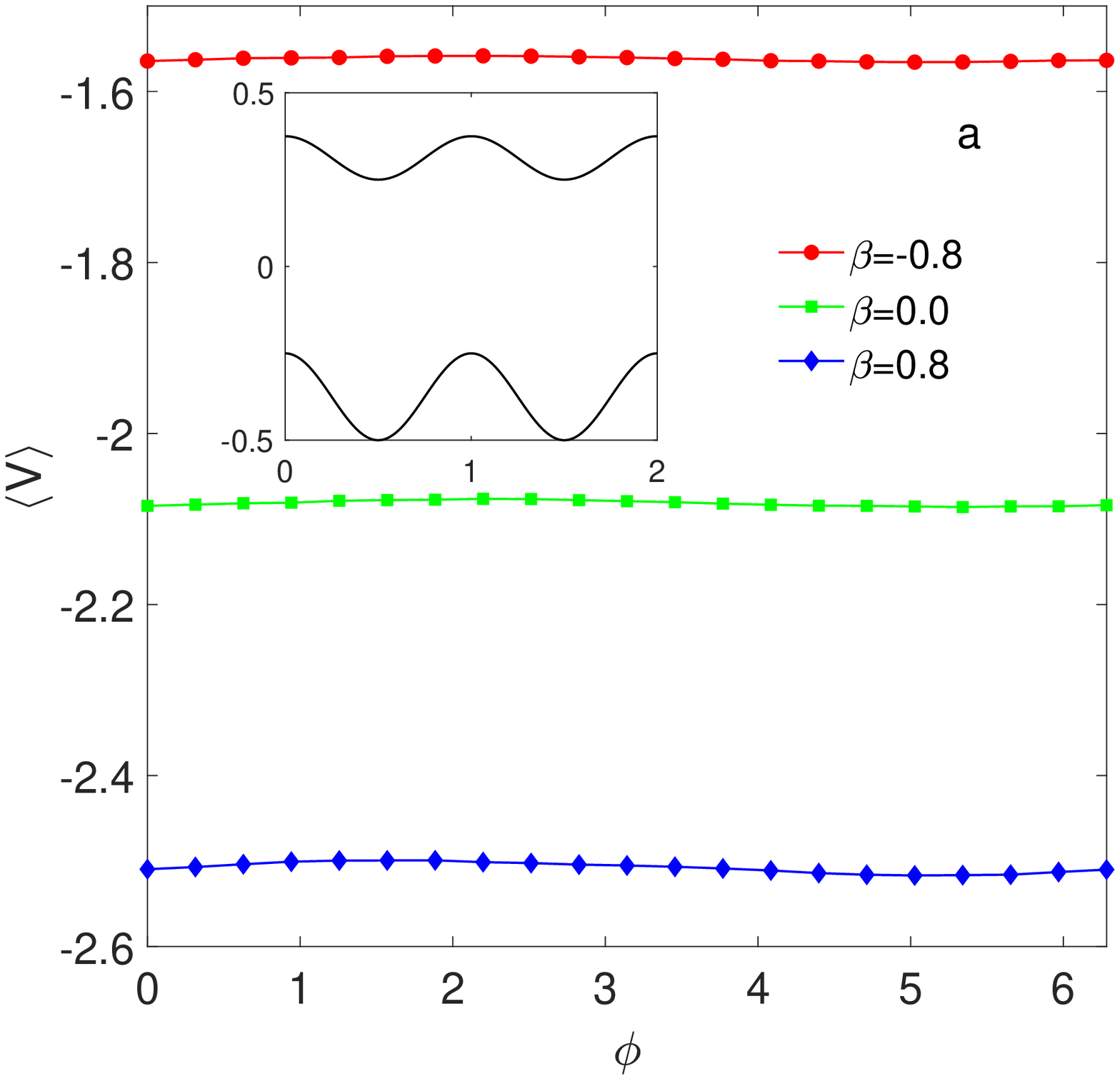}
}
\subfigure{
\includegraphics[height=6cm,width=7cm]{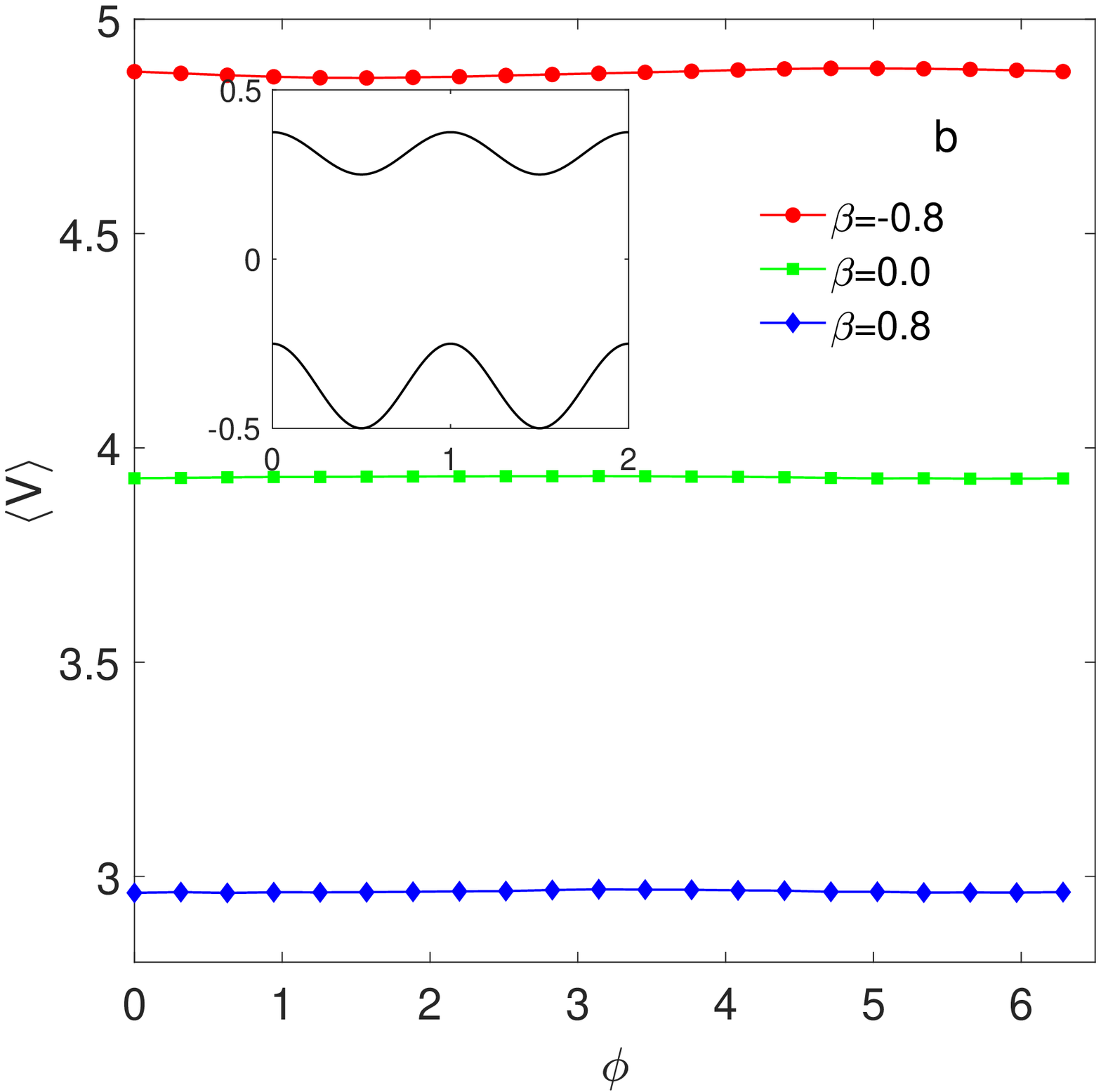}
}
\caption{The average velocity $\langle V\rangle$ as a function of $\phi$ with different asymmetry parameter $\beta$. The other parameters are $\epsilon=0.5$, $\Delta=0.5$, $\alpha=1.2$, $\sigma=0.5$, $v_0=0.1$, $\omega=0.2$, $x_L=y_L=1.0$, $Q_\theta=0.2$, $\tau_{\theta}=1.0$:(a)$\mu=-1.0$, (b)$\mu=1.0$.}
\label{VPhi}
\end{figure}

\begin{figure}
\centering
\subfigure{
\includegraphics[height=6cm,width=7cm]{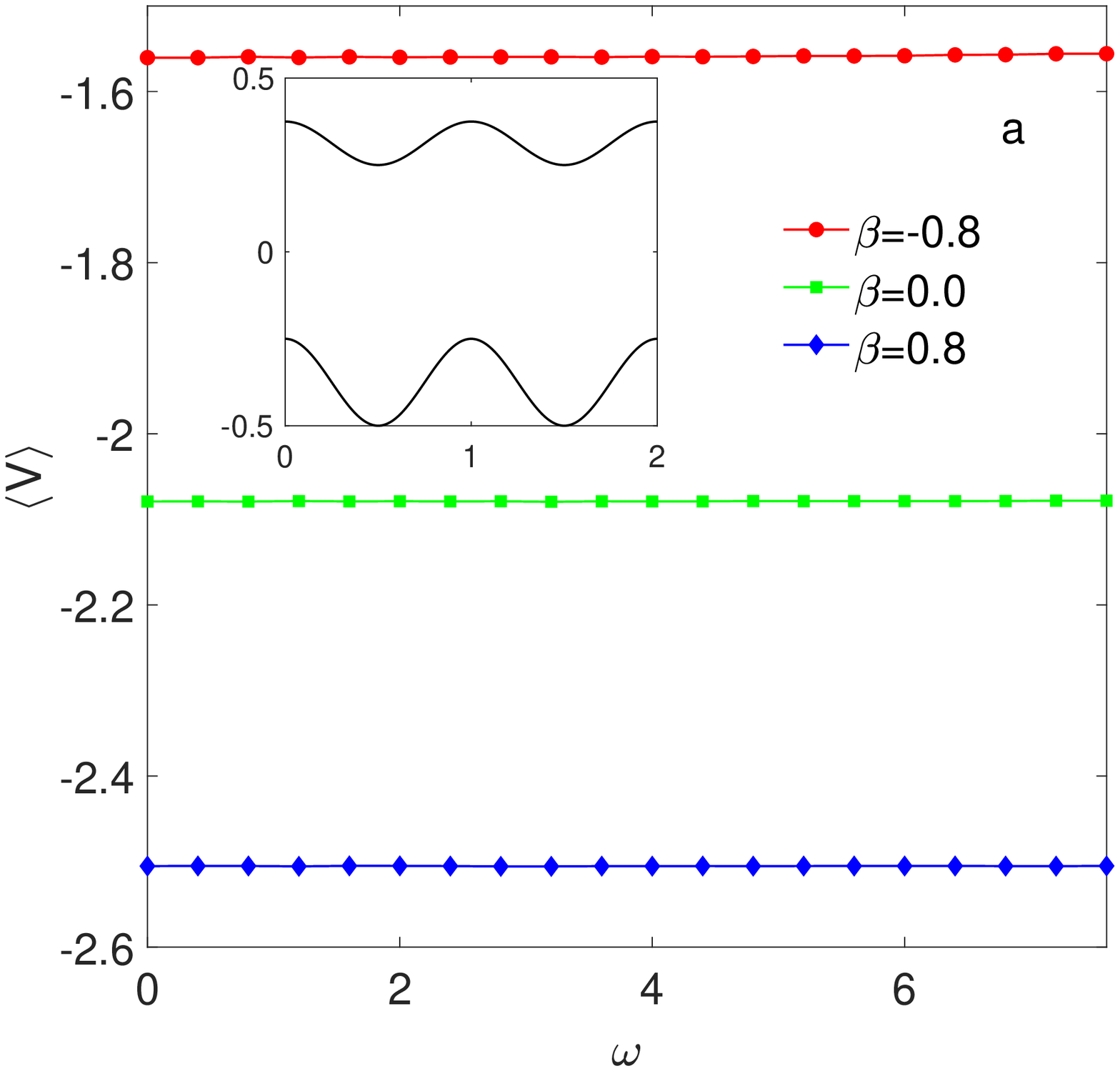}
}
\subfigure{
\includegraphics[height=6cm,width=7cm]{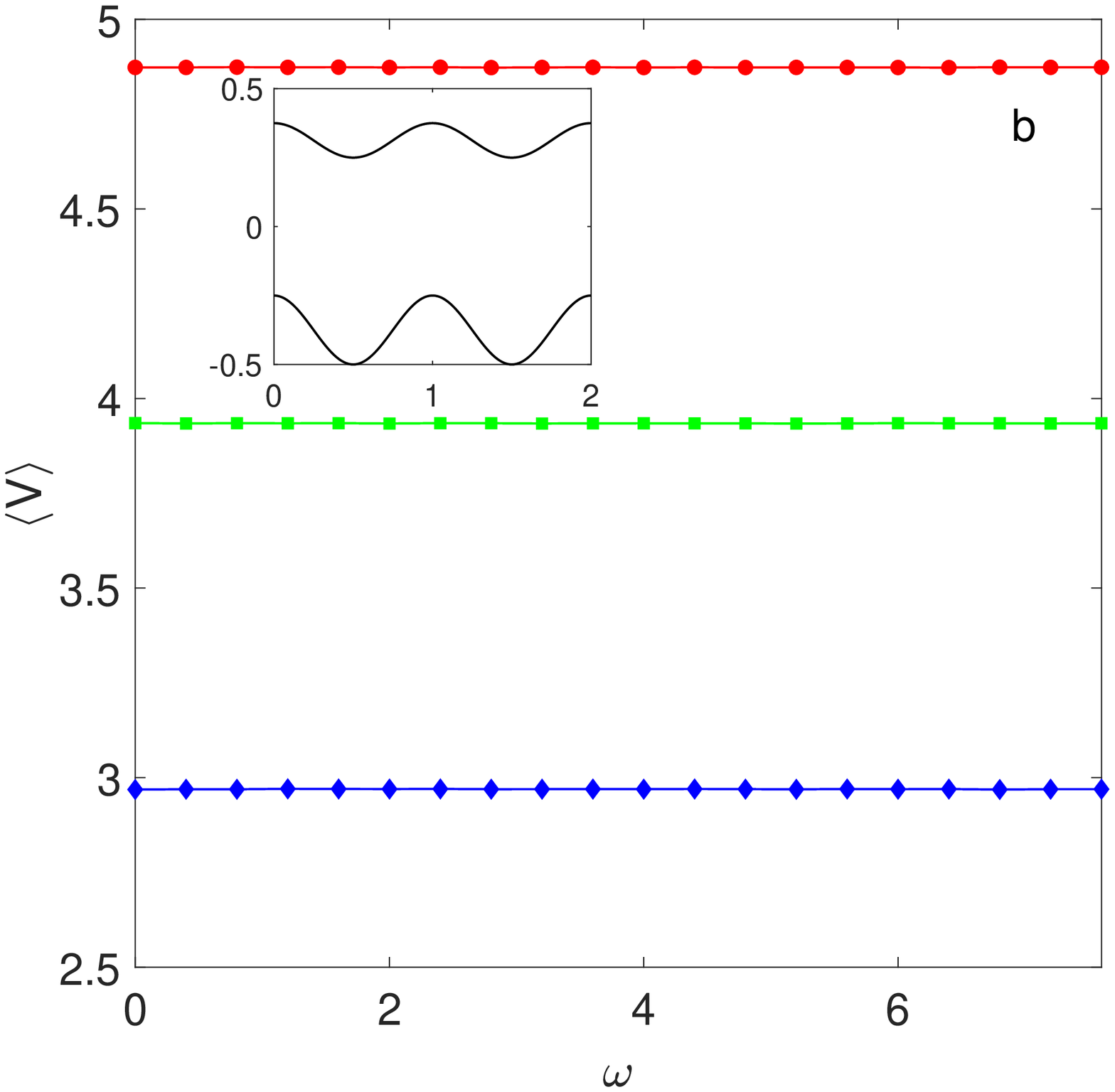}
}
\caption{The average velocity $\langle V\rangle$ as a function of angular velocity $\omega$ with different asymmetry parameter $\beta$. The other parameters are $\epsilon=0.5$, $\Delta=0.5$, $\phi=\pi$, $\alpha=1.2$, $\sigma=0.5$, $v_0=0.1$, $x_L=y_L=1.0$, $Q_\theta=0.2$, $\tau_{\theta}=1.0$:(a)$\mu=-1.0$, (b)$\mu=1.0$.}
\label{Vomega}
\end{figure}

Fig. \ref{VPhi} and Fig. \ref{Vomega} show $\langle V\rangle$ as functions of $\phi$($\phi$ is for tuning the shift of the upper wall from corresponding position of lower wall) and the angular velocity $\omega$. We find $\langle V\rangle$ almost remain unchanged with increasing $\phi$ and $\omega$, this means changes of $\phi$ or $\omega$ have no effect on the directional transport.

\begin{figure}
\centering
\subfigure{
\includegraphics[height=6cm,width=7cm]{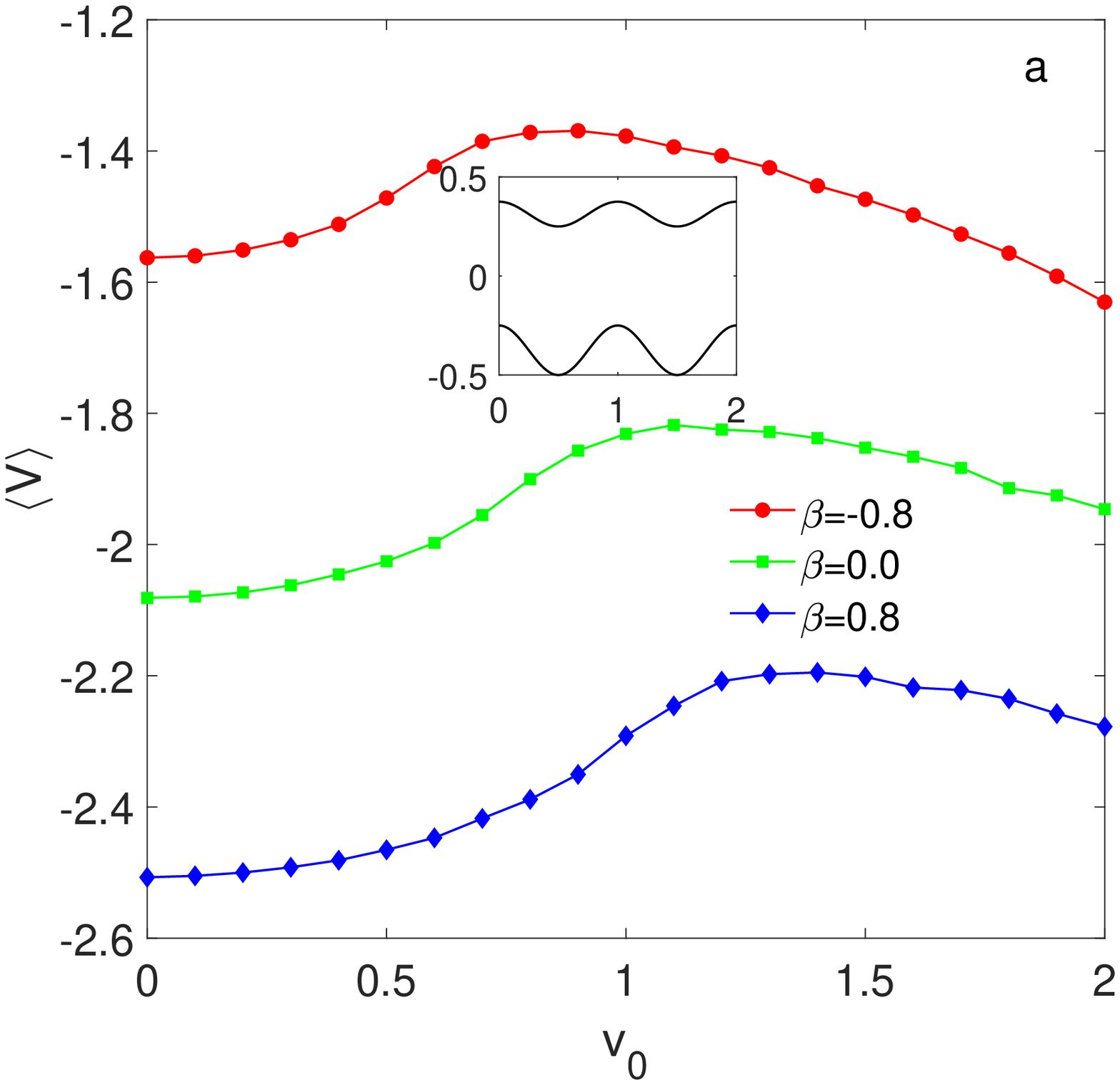}
}
\subfigure{
\includegraphics[height=6cm,width=7cm]{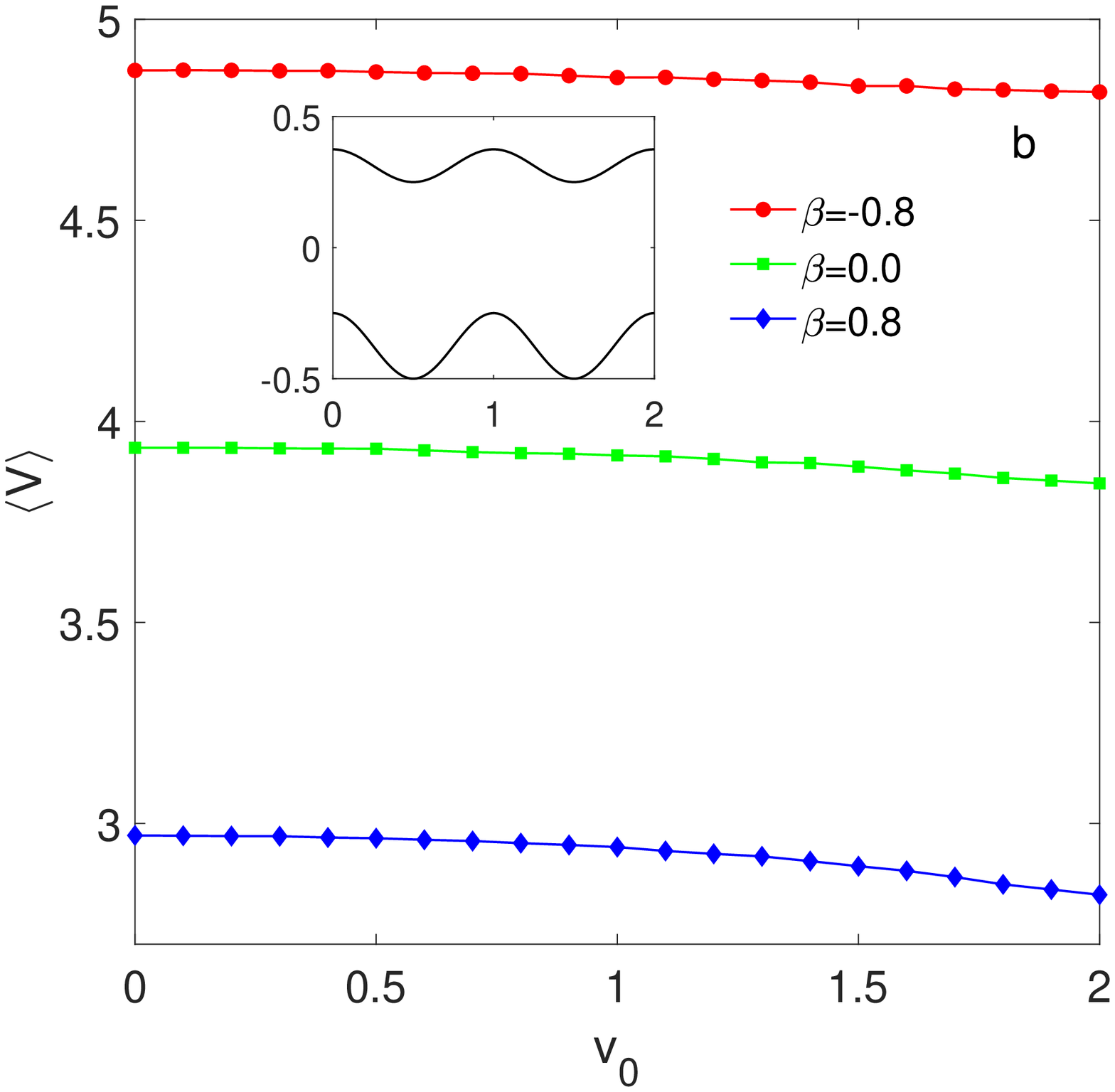}
}
\caption{The average velocity $\langle V\rangle$ as a function of $v_0$ with different asymmetry parameter $\beta$. The other parameters are $\epsilon=0.5$, $\Delta=0.5$, $\phi=\pi$,  $\alpha=1.2$, $\sigma=0.5$, $\omega=0.2$, $x_L=y_L=1.0$, $Q_\theta=0.2$, $\tau_{\theta}=1.0$:(a)$\mu=-1.0$, (b)$\mu=1.0$.}
\label{Vv0}
\end{figure}

The dependence of $\langle V\rangle$ on the self-propelled speed $v_0$ with different $\beta$ is shown in Fig.\ref{Vv0}. In Fig.\ref{Vv0}(a)($\mu=-1.0$), we find $\langle V\rangle<0$ and $\langle V\rangle$ has a maximum with increasing $v_0$, so proper self-propelled speed will inhibit the particle directional transport in $-x$ direction. $\langle V\rangle\neq0$ when $v_0=0$, this means inert particle confined in channel will move in $-x$ direction. In Fig.\ref{Vv0}(b)($\mu=1.0$), $\langle V\rangle$ decreases with increasing self-propelled speed $v_0$. So large $v_0$ will inhibit the particle moving in $x$ direction.

\begin{figure}
\centering
\subfigure{
\includegraphics[height=6cm,width=7cm]{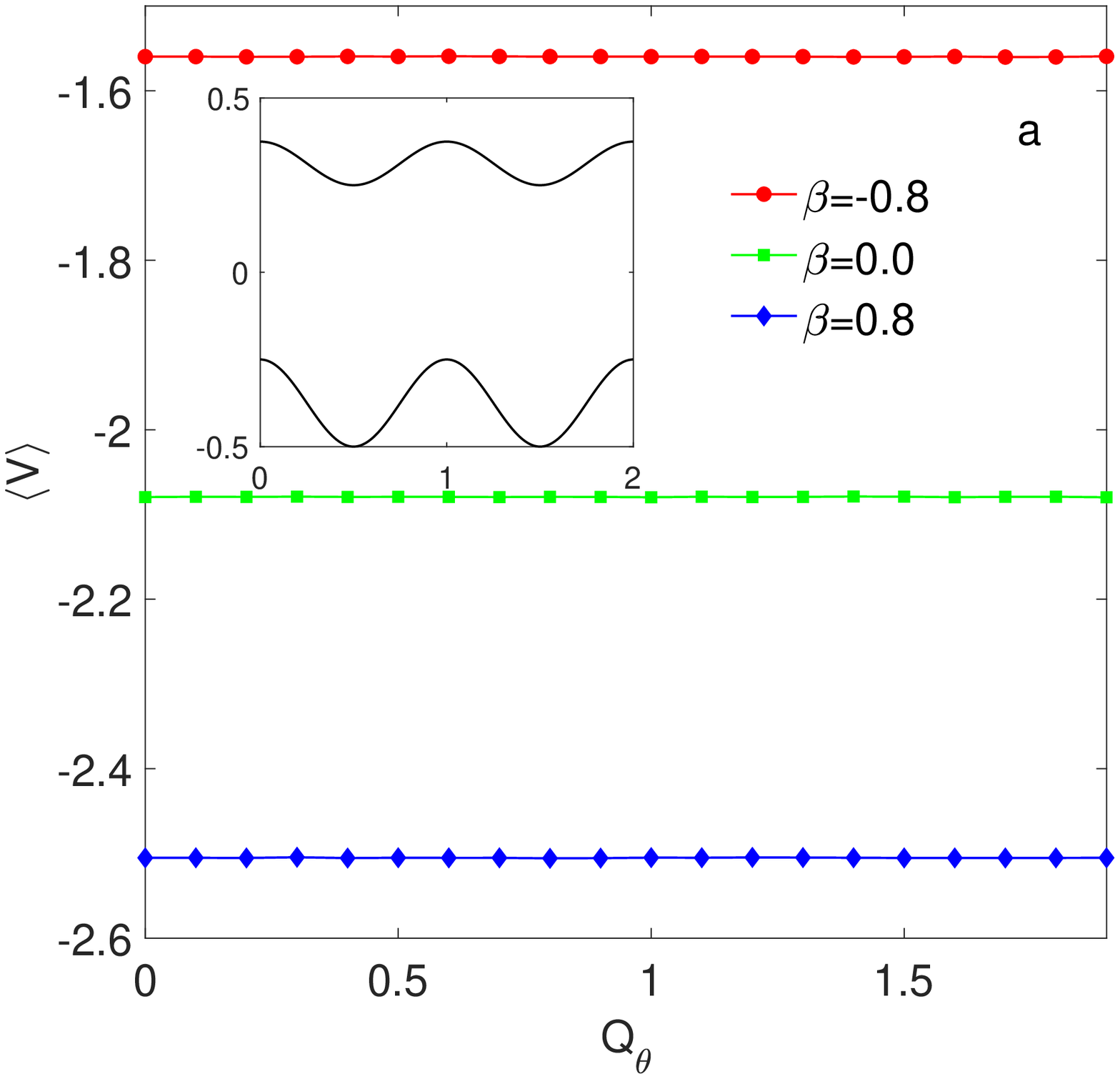}
}
\subfigure{
\includegraphics[height=6cm,width=7cm]{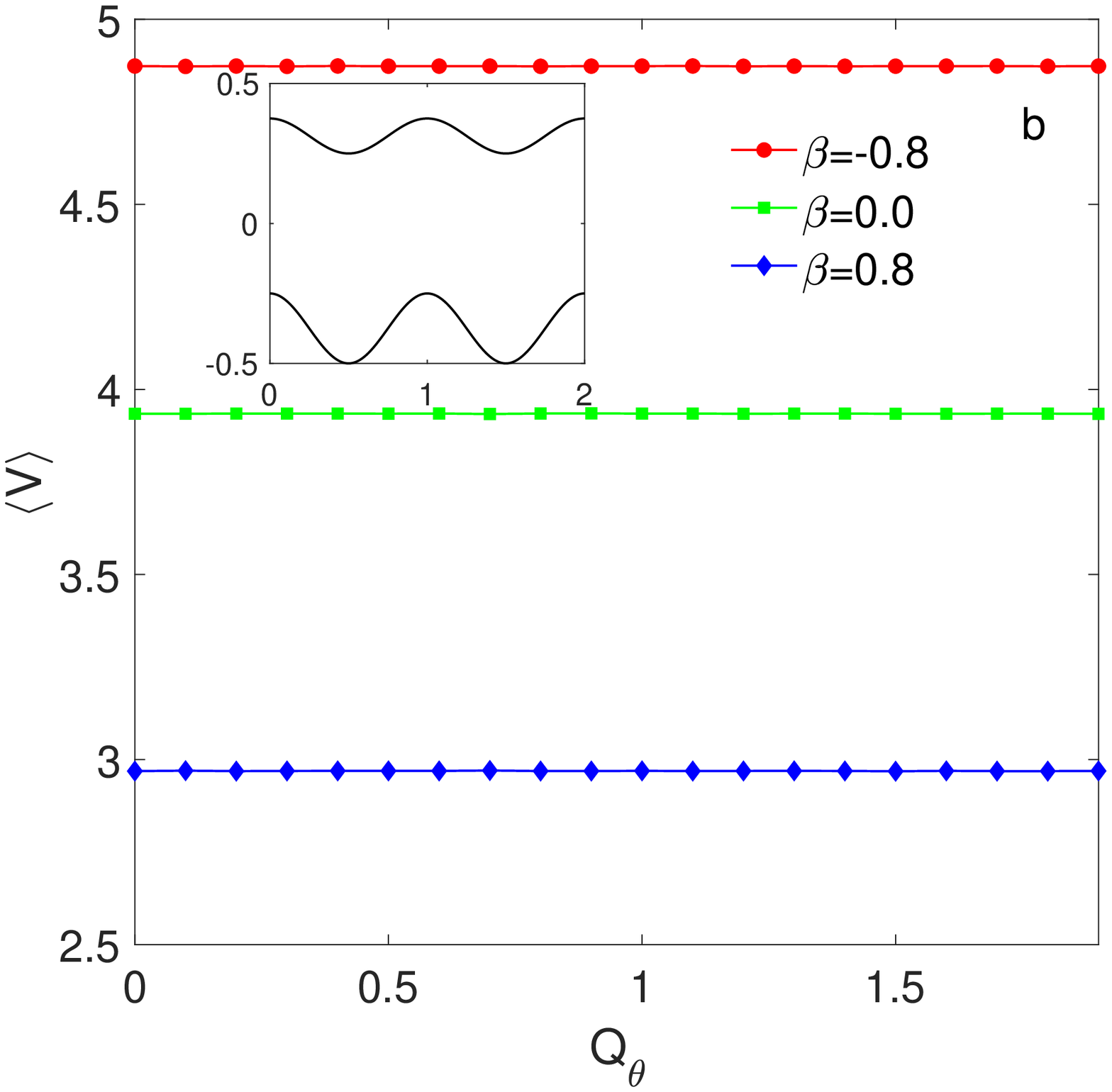}
}
\caption{The average velocity $\langle V\rangle$ as a function of noise intensity $Q_\theta$ with different asymmetry parameter $\beta$. The other parameters are $\epsilon=0.5$, $\Delta=0.5$, $\phi=\pi$, $\alpha=1.2$, $\sigma=0.5$, $v_0=0.1$, $\omega=0.2$, $x_L=y_L=1.0$, $\tau_{\theta}=1.0$:(a)$\mu=-1.0$, (b)$\mu=1.0$.}
\label{VQtheta}
\end{figure}

\begin{figure}
\centering
\subfigure{
\includegraphics[height=6cm,width=7cm]{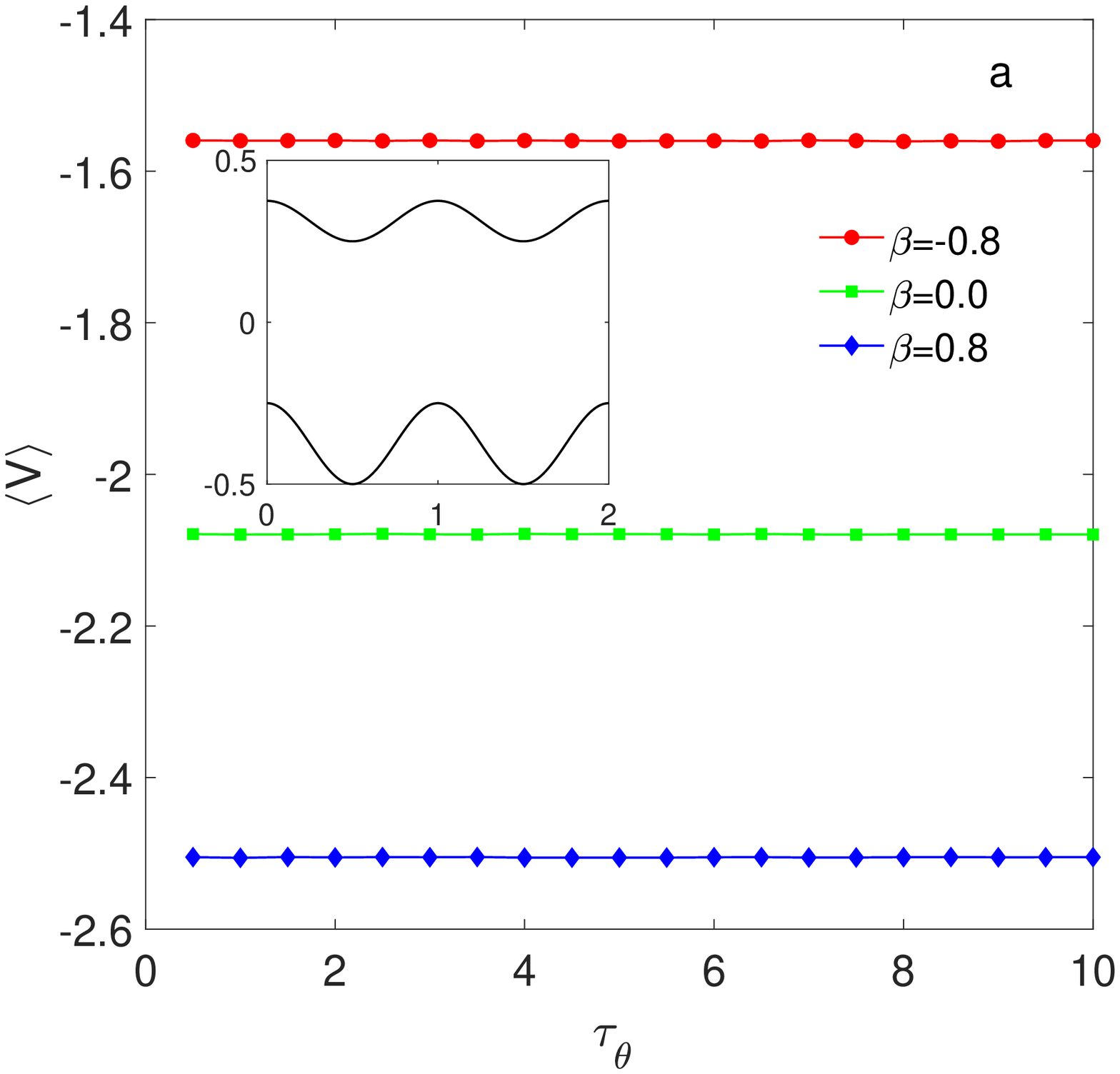}
}
\subfigure{
\includegraphics[height=6cm,width=7cm]{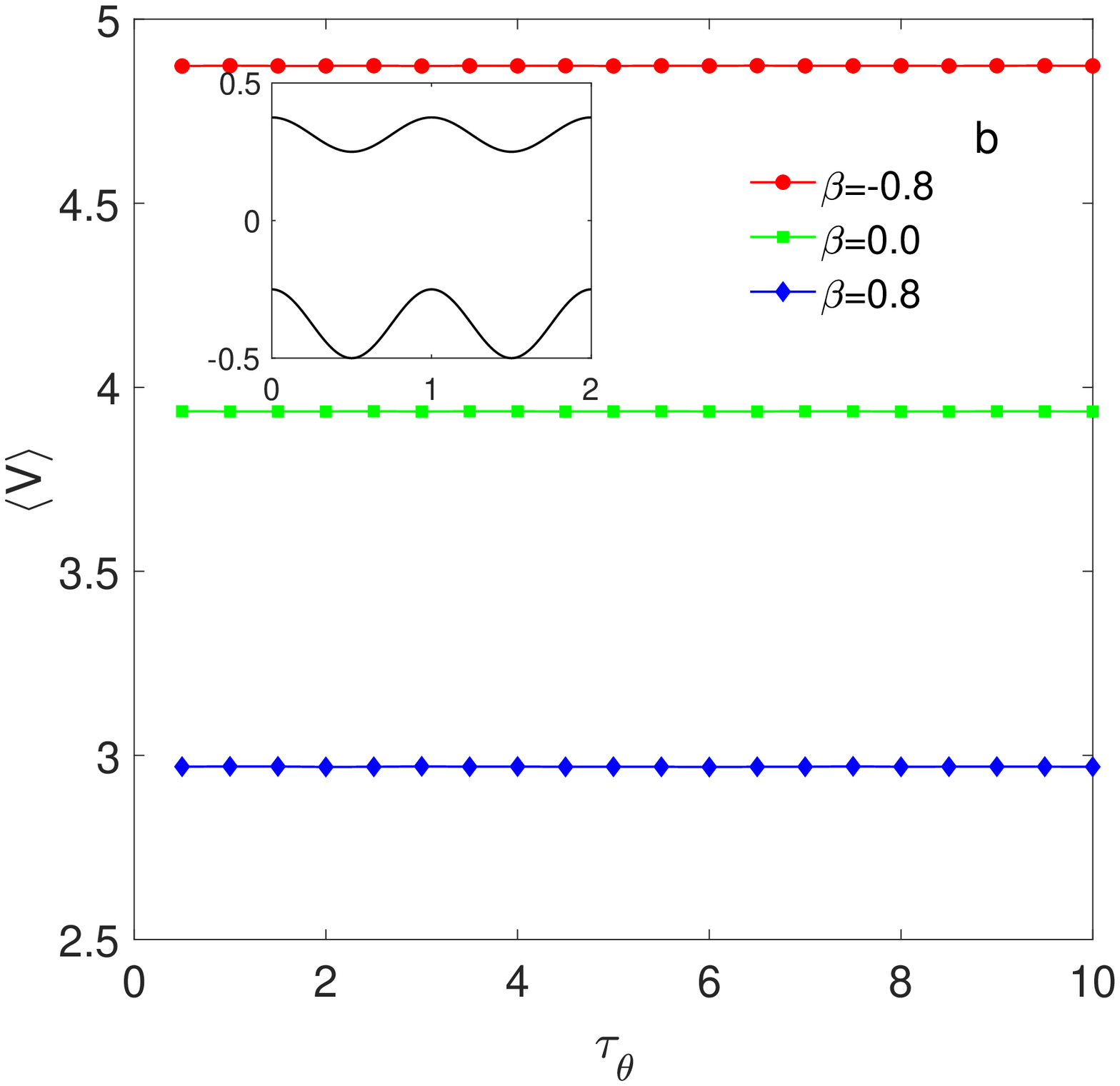}
}
\caption{The average velocity $\langle V\rangle$ as a function of self-correlation time $\tau_\theta$ with different asymmetry parameter $\beta$. The other parameters are $\epsilon=0.5$, $\Delta=0.5$, $\phi=\pi$, $\alpha=1.2$, $\sigma=0.5$, $v_0=0.1$, $\omega=0.2$, $x_L=y_L=1.0$, $Q_\theta=0.2$, $\tau_{\theta}=1.0$:(a)$\mu=-1.0$, (b)$\mu=1.0$.}
\label{VTautheta}
\end{figure}
The average velocity $\langle V\rangle$ as functions of of noise intensity $Q_{\theta}$ and self-correlation time $\tau_\theta$ with different $\beta$ is reported in Figs.\ref{VQtheta} and \ref{VTautheta}. We find there is almost no change for $\langle V\rangle$ with increasing $Q_{\theta}$ or $\tau_\theta$, so angle Gaussian noise has little effect on the particle transport. 

\section{\label{label4}Conclusions}
In this paper, we numerically studied the transport phenomenon of self-propelled particle confined in corrugated channel with L\'{e}vy Noise. The parameters of L\'{e}vy noise, i.e. the stability index, the asymmetry parameter, the scale parameter, the location parameter and the parameters of confined corrugated channel have joint effects on the particle. There exit flow reverse phenomena with increasing mean parameter. $\langle V\rangle$ shows complex behavior with increasing stability index. The $\langle V\rangle-\Delta$ curve forms the shape of $W$ when the distribution is skewed to the left. The $\langle V\rangle-\Delta$ curve forms the shape of $M$ when the distribution is skewed to the right. Angle Gaussian noise has little effect on the particle transport.
\section{Acknowledgments}

Project supported by Natural Science Foundation of Anhui Province(Grant No:1408085QA11) and College Physics Teaching Team of Anhui Province(Grant No:2019jxtd046).

\section{Referencing}

\end{document}